\documentclass{article}

\usepackage[margin=1in]{geometry}
\usepackage{ hyperref,amsthm,enumerate,
            imakeidx, xparse, mathtools,
            xcolor, amssymb, tensor, 
            soul, graphicx ,titlesec,  appendix, tikz,
            amsmath,scalerel, comment, float, makecell} 
\usepackage[most]{tcolorbox}
\usepackage{stackengine}
\usetikzlibrary{arrows}
\newcommand{\G}{\mathcal{G}}
\newcommand{\A}{\mathcal{A}}
\newcommand{\F}{\mathcal{F}}
\newcommand{\B}{\mathcal{B}}
\newcommand{\M}{\mathcal{M}}
\newcommand{\cH}{\mathcal{H}}
\newcommand{\K}{\mathcal{K}}
\newcommand{\Q}{\mathcal{Q}}
\newcommand{\J}{\mathcal{J}}

\newcommand{\g}{\mathfrak{g}}
\newcommand{\fd}{\mathfrak{d}}

 \renewcommand{\so}{\mathfrak{so}}
\newcommand{\iso}{\mathfrak{iso}}

\newcommand{\R}{\mathbb{R}}

\newcommand{\SO}{\mathrm{SO}}
\newcommand{\ISO}{\mathrm{ISO}}

\newcommand{\semir}{>\!\!\!\lhd}
\newcommand{\semil}{\rhd\!\!\!<}

\newcommand{\be}{\begin{equation}}
\newcommand{\ee}{\end{equation}}
\newcommand{\beq}{\begin{equation}}
\newcommand{\eeq}{\end{equation}}
\newcommand{\bes}{\begin{eqnarray}}
\newcommand{\ees}{\end{eqnarray}}

\newcommand{\cA}{{\cal A}}
\newcommand{\cD}{{\cal D}}
\newcommand{\cE}{{\cal E}}
\newcommand{\cF}{{\cal F}}

\newcommand{\cT}{{\cal T}}

\newcommand{\cB}{{\cal B}}
\newcommand{\cJ}{{\cal J}}

\newcommand{\cP}{{\cal P}}
\newcommand{\cZ}{{\cal Z}}
\newcommand{\cR}{{\cal R}}
\newcommand{\cL}{{\cal L}}
  
\def\demi{{\frac{1}{2}}}

\def\dr{{\rightarrow}}

\newtheorem{proposition}{Proposition}

\hypersetup{
pdfstartview = {FitH},
}
\hypersetup{
	colorlinks=true,         
	linkcolor=purple,          
	citecolor=red,        
	urlcolor=blue            
}

\newcommand{\la}{\langle}
\newcommand{\ra}{\rangle}

\def\dr{{\,\rightarrow\,}}
\def\mone{^{-1}}

\usepackage{authblk}
  \allowdisplaybreaks
\title{Discretization of 4d Poincar\'e $BF$ theory: from groups to 2-groups }

\begin{document}

\author[1]{\sffamily Florian Girelli\thanks{florian.girelli@uwaterloo.ca}}
\author[1]{\sffamily Panagiotis Tsimiklis\thanks{ptsimiklis@uwaterloo.ca}}

\affil[1]{\small Department of Applied Mathematics, University of Waterloo, 200 University Avenue West, Waterloo, Ontario, Canada, N2L 3G1}

\maketitle
\begin{abstract}
      We study the discretization of a Poincar\'e/Euclidean BF theory. Upon the addition of a boundary term this theory is equivalent to the  BFCG theory defined in terms of the Poincar\'e/Euclidean 2-group.  At an intermediate step in the discretization, we note that there are multiple options for how to proceed. One  option brings us back to recovering the discrete variables and phase space of BF theory. Another option allows us to rediscover the phase space related to the G-networks given in \cite{Asante:2019lki}.  Indeed,  our main result is that we are now able to relate the continuum fields with the discrete variables in \cite{Asante:2019lki}.  This relation is important to determine how to implement the simplicity constraints to recover gravity using the BFCG action. In fact we show that such relation is not as simple as in the BF discretization: the discretized variable on the triangles actually depend on several of the continuum fields instead of solely the continuum $B$-field. 
      We also compare and contrast the discretized BF and BFCG models as pairs of of dual 2-groups. 
      This work highlights (again) how the choice of boundary term influences the resulting symmetry structure of the \textit{discretized theory} -- and hence ultimately the choice of quantum states.
\end{abstract}
\tableofcontents

\section{Introduction}
Recent developments have pointed out the importance of boundary terms in the construction of a quantum theory of gravity \cite{Freidel:2020ayo, Freidel:2020svx, Freidel:2020xyx, Dupuis:2019unm}. Indeed we expect the quantum states of geometry to be given as representations of the symmetries at hand. The symmetries depend on the choice of variables used, as can be seen when comparing the metric and Palatini formalism. What was under-appreciated before is how adding boundary terms to the action might  greatly influence the   resulting symmetries  relevant to the construction of quantum states. 

We illustrate this point by considering the 4d ``Euclidean" BF theory, i.e.\ a BF theory where the underlying gauge group is the Euclidean (or similarly the Poincar\'e) group. Depending on whether or not we introduce a (very simple) boundary term to the theory, the symmetries of the quantum theory are given by two different types of (strict) Lie 2-groups --- a categorified version of a group, also called  a crossed module \cite{Baez:2010ya}. From a pragmatic point of view, we can understand the difference between a group and a 2-group in the following way. If elements of a (1-)group $G$ can be naturally associated to paths (holonomies), then elements in a 2-group are elements of a pair of groups $G$ and $H$ --- with some additional maps between them --- which decorate  paths and faces. It is widely expected that such structure (or their deformation) is the right structure to probe 4d topology just like (quantum) groups are the relevant structure to probe 3d topology \cite{Baez:1995xq}. 

4d \textit{BF} theory for a $d$ dimensional Lie group $G$ with Lie algebra $\g$ is a theory which can be seen as a 2-gauge theory, i.e.\ a gauge theory built on a 2-group \cite{Baez:2010ya, Girelli:2003ev}. However upon discretization, it is well known that we recover a copy of $T^*G\cong G \ltimes \g^* \cong G \ltimes \R^d$ associated to each link/triangle pair.  
The usual quantum states, called spin networks, are built from representations of $G$ 
so that in the end there is not much left of the 2-group picture. Indeed, a 1-group can be seen as a trivial 2-group. 

Following the key insights by Dittrich and Geiller \cite{Dittrich:2014wda, Bahr:2015bra, Delcamp:2018sef}, it was realized that we could instead build the quantum theory for a \textit{BF} theory from representations of the (quantum) group $\R^d$ (which is still a trivial 2-group). This amounts to a change of polarization \cite{Dupuis:2017otn}, which can actually be obtained by the  addition of a simple boundary term to the initial action. This new representation pioneered by Dittrich and Geiller is important as it allows us to discuss quantum states of geometry defined in terms of the frame field (as opposed to functions of the connection) in the quantum gravity context.        

Most approaches to construct a 4d quantum gravity model rely on using a 4d \textit{BF} theory \cite{rovelli2004}. The usual one consists of taking the underlying group to be the Lorentz group $\SO(1,3)$, however, 
it was shown that one could also use the Poincar\'e group $\ISO(1,3)$ \cite{Mikovic:2011si, Mikovic:2015hza, Mikovic:2018vku, Belov:2018uko}. In either case, gravity can be recovered by imposing the simplicity constraints on the 2-form $B$.  The new interesting part in the Poincar\'e formulation is that the translational part is associated to the frame field. Hence the frame field is  already present at the BF level, contrary to the Lorentz formulation where it appears only through the simplicity constraints\footnote{In fact in the Lorentz formulation, one  really has two intrinsic frames and the simplicity constraint  identifies them \cite{Freidel:2020ayo}.}.   

Interestingly, Poincar\'e \textit{BF} theory can be seen to be equivalent --- up to a boundary term --- to another topological theory based on a non-trivial 2-group, the Poincar\'e 2-group \cite{Mikovic:2011si, Mikovic:2015hza, Mikovic:2018vku}. This boundary term now implies only a partial change of polarization.
It is only in the translational that the configuration and configuration variables are swapped with respect to the \textit{BF} formulation. In a sense, we only do half of the dualization done by Dittrich and Geiller\footnote{We will comment in the discussion section on the notion of "semi-dualization" introduced by Majid which seems to be at play here.}. The new action is called the \textit{BFCG} action \cite{Girelli:2007tt} and was introduced as the continuum counterpart of the topological invariant built on 2-groups, the Yetter model \cite{yetter1993tqft}. In terms of actions, the symmetries are of the Poincar\'e \textit{BF} theory or the \textit{BFCG} theory are obviously  the same, they are just packaged differently since the what is called configuration or momentum differs between the two actions. This is what will create the difference at the discrete/quantum level   between the two theories.  

Since 2-groups (or 2-categories) seem to be a natural tool with which to probe  the topology of 4d spaces \cite{Baez:1995xq}, it is expected that they should be useful in building a quantum theory for gravity. With this in mind, a partition function has been constructed based on the representation theory of the 2-Euclidean group, the KBF model \cite{Baratin:2014era, Korepanov:2002tr, Korepanov:2002tp, Korepanov:2002tq}. It was recently shown how such model could  actually be related to the partition function of the \textit{BFCG} action for the 2-Euclidean group \cite{Asante:2019lki}. Part of the argument relied on the proposal (by B. Dittrich  \cite{Asante:2019lki}) of a phase space structure for the polyhedron which seemed to be naturally related to the \textit{BFCG} variables. This discretization was based on an educated guess and the correspondence with the continuum variables was not identified. \textit{One of the key results of this paper is to provide such relation.} This relation is especially important to build the quantum gravity amplitude. Indeed, it is naturally expected that the discretized \textit{B} field should be constrained by the discrete simplicity constraint to obtain the quantum gravity amplitude. However doing this led to some possibly overly simplified model \cite{Mikovic:2015hza, Mikovic:2018vku}. Imposing the simplicity constraint on the discretized \textit{B} field is justified provided that the discretized \textit{B} field is only dependent on the continuum $B$ field.
\textit{We will show here that the discretized \textit{B} field is actually a function not only of the \textit{B} field but also of some other fields.} This implies that implementing a naive simplicity constraint on such   discretized \textit{B} field is probably not the right thing to do to recover the gravity regime.           

\medskip

The discretization procedure first introduced  in \cite{Freidel:2011ue}  has already shown its power in several instances. The first step is to decompose the manifold of interest into cells. The next step is to determine what happens to the degrees of freedom in these cells. We use a \textit{truncation}: in the bulk we only consider the fields on-shell. At least for the cases of interest,  we can then express the symplectic data solely in terms of fields living on the boundary.  By gluing the cells together (while imposing continuity between neighbouring cells), we obtain a way to explicitly evaluate the symplectic potential and hence identify both the discretized variables and their phase space structure. In particular we recover, in the \textit{BF} case based on gauge group $G$, the usual $T^*G$ phase space, associated to each link of the dual complex. This approach allows one to recover the particle degrees of freedom in the 3d gravity context \cite{Freidel:2018pbr} or the quantum group symmetry for 3d gravity with a cosmological constant \cite{Dupuis:2020ndx}. Another interesting outcome of this discretization is the emphasis that a choice of polarization must be made \cite{Dupuis:2017otn, Shoshany:2019ymo}. In particular there is always the choice on where to discretize the different variables. While at the continuum level this might seem benign, at the quantum level it is important. Indeed, at the discrete level this choice restricts the set of symmetries that are imposed first. For example in the \textit{BF} case, we can have the Lorentz symmetries or the translational symmetries, as we alluded before. The polarisation choice in that case affects which of these two symmetries is imposed on the spin networks, and which arise in the dynamics.

\medskip

In present work, we show how this discretization approach allows us to recover the discrete phase space proposed in \cite{Asante:2019lki} starting from the \textit{BFCG} action. The main outcome is the expression of the discretized variables in terms of the (smeared) continuum fields which arise in the discreization procedure. 
Another outcome we find interesting is that the nature of the symmetries at play, which is very important for the construction of the quantum states, is very much dependent on the  type of boundary terms one considers. While we discretize \textit{BF}, a very well known theory, we are able to find non-trivial 2-group symmetries. We discussed above how changing the polarization was impacting the symmetry structure, dealing either with the gauge symmetries or the translational symmetries. Now with a partial change of polarization we will have either some standard group symmetries (the aforementioned gauge/translational symmetries) or a non trivial 2-group symmetry.

\paragraph{Outline} In section \ref{bftheory}, we review the basics of constructing a BF theory and its symmetries. We then go on to show how to get \textit{BFCG} theory from \textit{BF} theory by choosing the underlying group to be $\ISO(4)$ and adding a boundary term. For both theories, we recall the symplectic structure and charge structure. 

In section \ref{sec:discBF}, we describe the discretization process allowing to recover  the standard \textit{BF} discretization, which can be reinterpreted as recovering spinning tops for the group $G$ interacting through conservation of angular momenta.

In section \ref{sec: discbfcg}, we  identify an alternate way of simplifying the symplectic potential which leads to a natural definition of variables defined on edges and triangles of the cellular decomposition as well as on faces and links of its dual. We use a specific triangulation of the 3-sphere in order to identify the constraints. 
We show also how to recover the standard \textit{BF} discretization, starting from the BFCG variables: in a sense we can ``cancel" the boundary term transforming \textit{BF} into \textit{BFCG} in the discretization process by a proper choice of polarization. 

In section \ref{sec: 2gp}, we discuss how the 2-gauge theory structure applies to the current context and how we have different 2-groups at hand, according to the choice of boundary term, or polarization.

\section{BF and BFCG Theory}\label{bftheory}

\subsection{4d BF Theory}

\paragraph{Action}
We consider the Lie group $\cD$ which has a pair of   $d$-dimensional  Lie subgroups $\G$ and $\G^*$, with $\G^*$ abelian, such that $$\cD\cong \G\ltimes \G^*\cong \G^*\rtimes \G.$$   We denote by $\g$ and $\g^*$ the Lie algebra of $\G$ and $\G^*$  with   generators $e_i$ and  $e^*_i$ for $i \in \{1 \dots d\}$. The 2d dimensional Lie algebra of $\cD$ is noted $\fd\cong \g\ltimes \g^*\cong \g^*\rtimes\g$, with Lie brackets 
\begin{align}
  &  [e_i,e_j] = f^k{}_{ij}e_k, \quad  [e^{*i},e^{*j}]=0. \label{eq: lie alg}\\
   & [e_i,e^{*j}] = f^j{}_{ki}e^{*k}, \label{eq: double}
\end{align}
where index summation notation is used here and for the rest of this paper. The Lie bracket structure can be translated into an action of $\g$ on $\g^*$.  
\begin{align}
    \fd\cong \g\ltimes \g^*,
\end{align}
with the  action given by 
\begin{align}
    e_i \triangleright e^{*j} = f^j{}_{ki}e^{*k},
\end{align}
hence the Lie bracket can be written as 
\begin{align}
    [e_i,e^{*j}] = e_i \triangleright e^{*j} 
\end{align}

There is a natural pairing between $\g$ and $\g^*$ which is compatible with the Lie algebra brackets of $\g$ and $\g^*$ that extends to $\fd$.
\begin{align}
  &  \la e_i,e_j\ra = \la e^{*i}, e^{*j }\ra =0, \quad  \la e_i, e^{*j} \ra = \delta^j_i, \label{duality}
\\
   & [e_i,e^{*j}] = \la [e_k,e_i], e^{*j} \ra e^{*k} .\label{eq: double 2}
\end{align}
This pairing gives the relation
\begin{align}
    \la a,[b,c] \ra = -\la [b,a], c\ra, \quad \forall a, b, c\in\fd. \label{eq: invariance}
   \end{align}
 $\fd$ is called the (Drinfeld) double.

\medskip 

Let $\B$ be a $\g^*$ valued $2$-form and let $\A$ be a $\g$ valued 1-form. The curvature of $\A$ is denoted $\F$ and is defined as
\begin{align}
    \F = d\A +\demi [\A\wedge\A].
\end{align}
The  BF theory on the four dimensional manifold $\M$ is defined by the action
\begin{align}
    S_\G \coloneqq \int_\M \la \B \wedge  \F\ra = \int_\M \B_i \wedge \F^i. \label{eq:bf action}
\end{align}

\paragraph{Symmetries}The symmetries of the action (\ref{eq:bf action}) are given by the  gauge transformation parametrized by  the group element $G\in\G$, and the translation/shift symmetries parametrized by a $\g^*$ valued $1$-form $\eta$.
\begin{align}
\textrm{Gauge transformations: } & \left\{\begin{array}{l}    \A \mapsto G^{-1} \A G +G^{-1}dG  \\ \B \mapsto G^{-1}\B G 
\end{array}\right.  \label{eq: gauge}\\
\textrm{Shift transformations: } & \left\{\begin{array}{l}   \A \mapsto \A \\ \B \mapsto \B +d_\A \eta 
\end{array}\right.  \label{eq: shift}
\end{align}
where $d_\A \eta = d\eta +[\A,\eta]$ (note that due to \eqref{eq: double},  $[\A,\eta]\in \g^*$ and so $d_A\eta \in \g^*$).
The action is invariant under the gauge transformation due to the invariance of $\la \cdot, \cdot \ra$ given in (\ref{eq: invariance}). 
 Under the shift transformation, the action is invariant up to a boundary term, 
\begin{align}
    S_\G \mapsto S_\G +\int _\M d\la \eta \wedge \F \ra,
\end{align}
where we used the Bianchi identity, $d_\A \F=0$. As we will recall in the next section, the shift symmetry can be interpreted as a 2-gauge transformation.

\paragraph{Equations of motion and potential} 
  The equations of motion associated with $S_\G$ are obtained by varying the  fields, $\A$ and $\B$:
\begin{align}
    \delta S_\G = \int_\M \la \delta B \wedge \F \ra - \int _\M \la d_\A \B\wedge \delta \A \ra +\int_\M d\la \B \wedge \delta \A  \ra\label{eq:variation}
\end{align}
The first two terms are the equations of motion:
\begin{align}
    \F = 0 \qquad d_\A \B=0 \label{eq: eom}
\end{align}
The third term in (\ref{eq:variation}) is the symplectic potential and is responsible for the Poisson structure. Denoting the boundary of $\M$ by $\partial \M \coloneqq M$, the potential is 
\begin{align}
    \Theta \coloneqq \int_M \la \B \wedge \delta \A  \ra.
\end{align}
The associated symplectic 2-form is
\begin{align}
    \Omega \coloneqq \delta \Theta = \int_M \la \delta \B \wedge \delta \A \ra.  
\end{align}
We note that the equations of motion imply two other sets of condition, namely 
\begin{align}
    \F = 0 \implies d_\A \F= 0 \qquad d_\A \B =0 \implies d_\A\big(d_\A \B\big)= [\cF,\B]=0 \label{eq: eom topo}
\end{align}
The first relation is  the standard Bianchi identity.  The condition  $[\cF,\B]=0$ is weaker than the flatness constraint. 

\paragraph{Charges/momentum maps. } We consider the infinitesimal versions of the gauge transformations and want to identify the charges generating such transformations. 
Let the infinitesimal version of (\ref{eq: gauge}) be parametrized by $\chi \in \g$,  and  the infinitesimal version of the shift transformation (\ref{eq: shift}) be parametrized by $\beta \in \g^*$,
\begin{align}
    \textrm{Infinitesimal gauge transformations: } & \left\{\begin{array}{l}  \delta _\chi \A = d_\A \chi  \\  \delta _\chi \B = [\B,\chi]\end{array}\right. \label{sym:BFg} \\
    \textrm{Infinitesimal shift transformations: } & \left\{\begin{array}{l} \delta _\beta \A =0   \\\delta _\beta \B = d_\A \beta\end{array}\right.  \label{sym:BFs}
\end{align}
We think of $\delta_\chi$ and $\delta_\beta$ as vector fields in the field space. We use notation such as $\delta_\chi \lrcorner \delta \A = \delta_\chi \A$ to express the interior product of a vector and a 1-form in field space. We define the charges as
\begin{align}
\delta \J_\chi &\coloneqq -\delta_\alpha \lrcorner \Omega =\int _M \la\delta \B\wedge \delta_\chi \A \ra -\int_M \la \delta_\chi\B \wedge \delta \A \ra\\
\delta \cP_\beta &\coloneqq -\delta _\beta \lrcorner \Omega = -\int_M \la \delta_\beta \B \wedge \delta \A \ra
\end{align}
Some manipulation reveals (provided we assume $\delta \alpha =0$ and $\delta \beta =0$, ie the parameters are not field dependent)
\begin{align}
    \J_\chi &= \int_M d\la \B \wedge \chi \ra-\int_M \la d_\A\B\wedge \chi\ra \approx \int_M d\la \B \wedge \chi \ra\\
    \cP_\beta &= -\int_M d\la \beta \wedge \A \ra -\int_M\la \beta \wedge \F\ra \approx -\int_M d\la \beta \wedge \A \ra,  
\end{align}
where $\approx$ means we went on-shell.  We note that the charges are essentially given by the corner charges specified by the variables $\B$ and $\A$, a 2-form and a 1-form respectively.  When the parameters are constant, we will call the charges \textit{global}.  We have used in each case the pull-back to $M$ of the equations of motion. These pull-backs are also interpreted as \textit{constraints}. A momentum map is a function on phase space generating the symmetry transformations. As such the constraints are  momentum maps. 
\begin{table}[h!]
    \centering
    \begin{tabular}{|c|c|c|}
    \hline
    Charge&    Momentum map & Symmetry\\\hline\hline
$\cB$&$d_\A\B$ & gauge transformation  \\
$\cA$&$\F$ & shift transformation\\
         \hline
    \end{tabular}
    \caption{Summarizing the charges, momentum maps and associated symmetries. }
    \label{tab:0}
\end{table}

\medskip

To anticipate a bit our results, we note that the global charges $\J$ and $\cP$ are simply generated by the configuration/momentum variables. This implies that where the phase space variables are discretized will directly influence how the charges will be discretized.   This means that we might get different types of symmetry structure (1-group versus 2-group) according to the choice of discretization of the variables.

\subsection{BFCG theory}
Let's revisit the action of BF theory and see how specifying the group might change our perspective.  We consider $\g$ to  be the Euclidian  (or Poincar\'e) Lie algebra  $\g=\iso(4)\cong   \so(4)\ltimes \R^4 $ and $\g^*$ is the dual abelian Lie algebra  $\g^*=\iso^*(4)\cong    \so^*(4)\times \R^{*4} $, with $\so^*(4)\cong \R^6$ and $\R^{*4}\cong \R^4$.
\begin{align}
\fd \sim \iso(4) \ltimes    \iso^*(4).
\end{align}
The subalgebra $\R^4$ is generated by $P^{\mu}$ and the rotation algebra $\so(4)$ is generated by $J^{\mu\nu}$. Greek indices range from 0 to 3. The Lie brackets are
\begin{align}
    [J^{\mu\nu},J^{\sigma\rho}] &= \eta^{\mu\rho}J^{\nu\sigma }+\eta^{\nu\sigma}J^{\mu\rho}-\eta^{\mu\sigma}J^{\nu\rho}-\eta^{\nu\rho}J^{\mu\sigma}\\
    [P^\mu,P^{\nu}]=0, &\quad 
    [J^{\mu\nu},P^\sigma]=\eta^{\mu\sigma}P^\nu-\eta^{\nu\sigma}P^\mu, \label{eq: algebra}
\end{align}
where $\eta$ is the flat metric. We will sometimes write $[J^{\mu\nu}, J^{\sigma\rho}] = f^{\mu\nu\sigma\rho}{}_{\alpha\beta}J^{\alpha\beta}$ where $f$ is the structure constant of $\so(4)$. The bilinear pairing is
\begin{align}
    \la P^*_\nu,P^\mu \ra =& \delta_\nu^\mu, \quad  
    \la J^*_{\mu\nu}, J^{\sigma\rho}\ra = \delta_\mu^\sigma\delta _\nu^\rho-\delta^\sigma_\nu\delta_\mu^\rho,
\end{align}
where the generators in the dual are distinguished by lowered indices and an asterix. The second term in the second pairing is to account for the antisymmetry of the indices. We will identify the subspace generated by $P^*$ as $\R^{4*}$, the subspace generated by $J^*$ as $\so(4)^*$, and  $\iso^*(4)\cong \R^{10}$.  
\begin{align}
    [P^*_\mu,P^*_\nu] = [J_{\mu\nu}^*,J_{\sigma\rho}^*]=[J_{\mu\nu}^*,P_\sigma^*] = [J_{\mu\nu}^*,P^\rho]=0.
\end{align}
The Lie brackets of the double is constructed according to\footnote{Writing \eqref{eq: double 2} in terms of $J$'s and $P$'s gives, for example, $[P^\mu,P^*_\nu] = \la [P^\sigma,P^\mu],P^*_\nu\ra P^*_\sigma+\demi \la [J^{\alpha\beta}, P^\mu],P^*_\nu \ra J^*_{\alpha\beta}$. }  (\ref{eq: double 2})
\begin{align}
    [P^\mu,P^*_\nu] =& \eta^{\mu\sigma}J^*_{\sigma\nu}\label{nontrivial}\\
    [J^{\sigma\rho},P^*_\nu] =& (\eta^{\rho\alpha}\delta_\nu^\sigma- \eta^{\sigma\alpha}\delta_\nu^\rho)P^*_\alpha\\
    [J^{\mu\nu},J^*_{\sigma\rho}]=& (\eta^{\alpha\nu}\delta_\rho^\mu-\eta^{\alpha\mu}\delta_\rho^\nu)J^*_{\alpha\sigma}+(\eta^{\alpha\mu}\delta_\sigma^\nu-\eta^{\alpha\nu}\delta^\mu_\sigma)J^*_{\alpha\rho}
\end{align}
The non-zero brackets $ [P^\mu,P^*_\nu]$ in \eqref {nontrivial} will play an important role in the discretization procedure. 
We decompose the fields $\A$ and $\B$ into their rotation and translation components:
\begin{align}
    \B &= B + \Sigma \qquad &B \in \so(4)^*\qquad \Sigma \in \R^{4*}\\
    \A &= A + C \qquad &A \in \so(4)\qquad C \in \R^4
\end{align}
The curvature is also rewritten in terms of its projections into subalgebras. 
\begin{align}
  \F = F +d_AC, \quad \textrm{ with }    F = dA +\demi[A\wedge A] \in \so(4) && d_AC = dC +[A\wedge C] \in \R^4.
\end{align}
Hence the action is
\begin{align}
    S_{ISO(4)} =& \int_\M \la B \wedge F \ra + \int_\M \la  \Sigma \wedge d_A C \ra = \int_\M \la B \wedge F \ra - \int_\M \la  d_A\Sigma \wedge  C \ra + \int_\M d\la \Sigma \wedge C \ra. 
\end{align}
The quantity $G \coloneqq d_A \Sigma$ is called the 2-curvature. The BFCG action is obtained from $S_{ISO(4)}$ up to a  boundary term \cite{Mikovic:2011si, Mikovic:2015hza},
\begin{align}
    S_{BFCG} =& \int_M \la B\wedge F \ra +\int_M \la C \wedge G \ra = S_{ISO(4)} - \int_\M d\la \Sigma \wedge C \ra.
\end{align}

\paragraph{Equations of motion and potential} The equations of motion can be determined by varying $S_{BFCG}$ with respect to $A$, $B$, $C$, and $\Sigma$. Alternatively, since a boundary term in the action does not change the equations of motion we can decompose the expression in (\ref{eq: eom}) into translation and rotation components. Either way, we get
\begin{align}
    F &=0 \qquad d_A C = 0\nonumber\\
    G &= 0 \qquad d_A B = -[C\wedge\Sigma]. \label{eq: bfcgeom}
\end{align}
We remind the reader that though $C$ and $\Sigma$ are both elements of a Lie subalgebra with trivial brackets, the bracket between them is given by \eqref{nontrivial} which  is not zero  but in $\so^*(4)$.
Similarly the potential is decomposed
\begin{align}
    \Theta_{ISO(4)} = \int_M \la B\wedge \delta A\ra +\int_M \la \Sigma \wedge \delta C \ra.
\end{align} 
The potential from $S_{BFCG}$ is slightly different,
\begin{align}
    \Theta_{BFCG} = \int_ M \la B \wedge \delta A \ra - \int _M \la C\wedge \delta \Sigma \ra.
\end{align}
The two potentials differ by a total functional derivative and therefore give the same 2-form\footnote{$\delta C \wedge \delta \Sigma = -\delta \Sigma \wedge \delta C$ since they are 1-forms in field space}:
\begin{align}
    \Omega = \int_M \la \delta B \wedge \delta A \ra +\int_M \la \delta \Sigma \wedge \delta C \ra.
\end{align}
We note that we also have the analogue of the Bianchi identity and its companion \eqref{eq: eom topo} which are still valid. Breaking them into components we have 
\begin{align}
\left.
\begin{array}{r} 
d_\A\F=0\\
d_A(d_A C)=0
\end{array}\right\}
&\rightarrow 
\left\{
\begin{array}{l}  
\, d_A F=0 \\ \, [F,C]=0
\end{array}
\right.
\\
   d_\A(d_\A\B)=[\F\wedge\B]=0
 &  \rightarrow 
\left\{\begin{array}{l}  \, [F,B]+ [d_AC\wedge\Sigma]=0 \\ 
 \, [F,\Sigma]=0
\end{array}\right.\label{continuous-edge-simp}
\end{align}

\paragraph{Symmetries} Now let's review the symmetries introduced in the previous section. We have now different names as the fields are interpreted in the 2-gauge theory picture \cite{Girelli:2007tt}.  
\begin{align}
    \textrm{Gauge : }     \left\{\begin{array}{l}  
    \delta _\chi \A = d_\A \chi  \\  \delta _\chi \B = [\B,\chi]
    \end{array}\right. 
    & \dr \begin{array}{l}   \chi= \alpha+ X \\ \alpha \in \so(4), \, X\in \R^4  \end{array}
    \dr 
     \left\{\begin{array}{l} 
     \textrm{1-gauge transformation }
     \left\{\begin{array}{l} 
  \delta _\alpha A = d_A\alpha \\ \delta _\alpha B = [B,\alpha]\\
    \delta _\alpha C = [C,\alpha] \\ \delta_\alpha \Sigma = [\Sigma,\alpha]     \end{array}\right.
     \\
    \textrm{2-shift } 
    \left\{\begin{array}{l}    
  \delta _X A=0  \\  \delta _X C= d_A X\\
     \delta _X B =[\Sigma , X] \\  \delta _X  \Sigma =0    \end{array}\right.
    \end{array}\right.
     \\
    \textrm{ Shift : }    \left\{\begin{array}{l} 
    \delta _\beta \A =0  \\
    \delta _\beta \B = d_\A \beta
    \end{array}\right. & \dr  \begin{array}{l}  \beta = \zeta +Y, \\ \zeta \in \R^{4*}, \, Y  \in \so(4)^*   \end{array}
     \dr  
     \left\{\begin{array}{l} 
     \textrm{2-gauge transformation }
     \left\{\begin{array}{l} 
     \delta_\zeta A =0\\ \delta_\zeta C =0 \\\delta _{\zeta} B = [C\wedge\zeta]\\ \delta _{\zeta} \Sigma = d_A \zeta 
     \end{array}\right.
     \\
    \textrm{1-shift } 
    \left\{\begin{array}{l}    \delta _Y A =0\\ \delta _Y C =0 \\\delta _{Y}B = d_AY \\ \delta _Y \Sigma = 0 
    \end{array}\right.
    \end{array}\right.
\end{align}

\paragraph{Charge and momentum maps. }
As before, we define the charges\footnote{A similar analysis was done by M. Geiller in some unpublished work.} 
\begin{align}
    \delta \cL_\alpha \coloneqq& - \delta_\alpha \lrcorner \Omega = -\int_M \la\delta_\alpha B\wedge \delta A \ra +\int_M \la\delta B \wedge \delta _\alpha A\ra - \int_M \la\delta _\alpha\Sigma \wedge \delta C\ra +\int_M \la\delta \Sigma \wedge \delta _\alpha C\ra \\
    \delta \cR_ {Y} \coloneqq& -\delta_{Y} \lrcorner \Omega = -\int_M \la\delta _{Y} B\wedge \delta A \ra\\
    \delta \K_{X} \coloneqq& -\delta _{X} \lrcorner \Omega = -\int_M \la\delta_X B \wedge \delta A\ra+\int_M \la\delta \Sigma \wedge \delta_X C\ra\\
    \delta \Q_{\zeta} \coloneqq& - \delta _{\zeta} \lrcorner \Omega = -\int_M \la\delta_{\zeta}B \wedge \delta A \ra-\int_M \la\delta _{\zeta}\Sigma \wedge \delta C\ra
\end{align}
After some algebra, we find that, still assuming a non-dependence of the parameters in terms of the fields, 
\begin{align}
    \cL_\alpha &= \int_M d\la B\wedge \alpha \ra - \int_M \la (d_A B + [C\wedge\Sigma])\wedge \alpha)\approx \int_M d\la B\wedge \alpha \ra\\
    \cR_{Y} &= -\int_M d\la Y \wedge A\ra - \int_M \la Y \wedge F \ra \approx -\int_M d\la Y \wedge A \ra\\
    \K_X &= \int_M d\la \Sigma \wedge X \ra - \int_M \la X\wedge d_A\Sigma \ra \approx \int_M d\la \Sigma \wedge X \ra \\
    \Q_{\zeta} &= -\int_M d\la \zeta \wedge C \ra -\int_M \la \zeta\wedge d_A C \ra \approx -\int_M d\la \zeta \wedge C \ra  
\end{align}
If the coefficients $\zeta,X,Y,\alpha$ are constant on the boundary, we deal then with ``global" charges. There is then no central extension in their Poisson algebra (see Appendix \ref{algebra}).   As before we have a set of constraints, which are the pull-back of the equations of motion to $M$. 

These constraints are momentum maps generating the symmetry transformations, we list them in table \ref{tab:mmas}. 
\begin{table}[h!]
    \centering
    \begin{tabular}{|c|c|c|}
    \hline
    Charge&    Momentum map & Symmetry\\\hline\hline
$B$&$d_A B + [C\wedge\Sigma]$  & 1-gauge transformation  \\
$\Sigma$&$d_\A\Sigma$ & 2-shift \\
$A$ &$F$ & 1-shift \\
$C$ &$d_A C$ & 2-gauge transformation\\
         \hline
    \end{tabular}
    \caption{Summarizing the charges, momentum maps and associated symmetries. }
    \label{tab:mmas}
\end{table}

\medskip

Once again anticipating, we can see that  that the discretized symmetries will different whether we consider the pair $(A,C)$ (ie $\cA$) discretized on the dual complex as would be done for regular BF theory (hence obtaining a 1-gauge theory), or if we consider the pair $(A,\Sigma)$     discretized on the dual complex as would be done for the BFCG theory (hence considering a 2-gauge theory). 

While the continuum theories are equivalent up to a boundary term, the discrete symmetry will be different.

\section{Discretization of 4d ISO(4) BF theory}\label{sec:discBF}

\subsection{Notation}

Let us set up the notations for the cellular decomposition once and for all.
\begin{figure}
    \centering
    \includegraphics[width=0.3\textwidth]{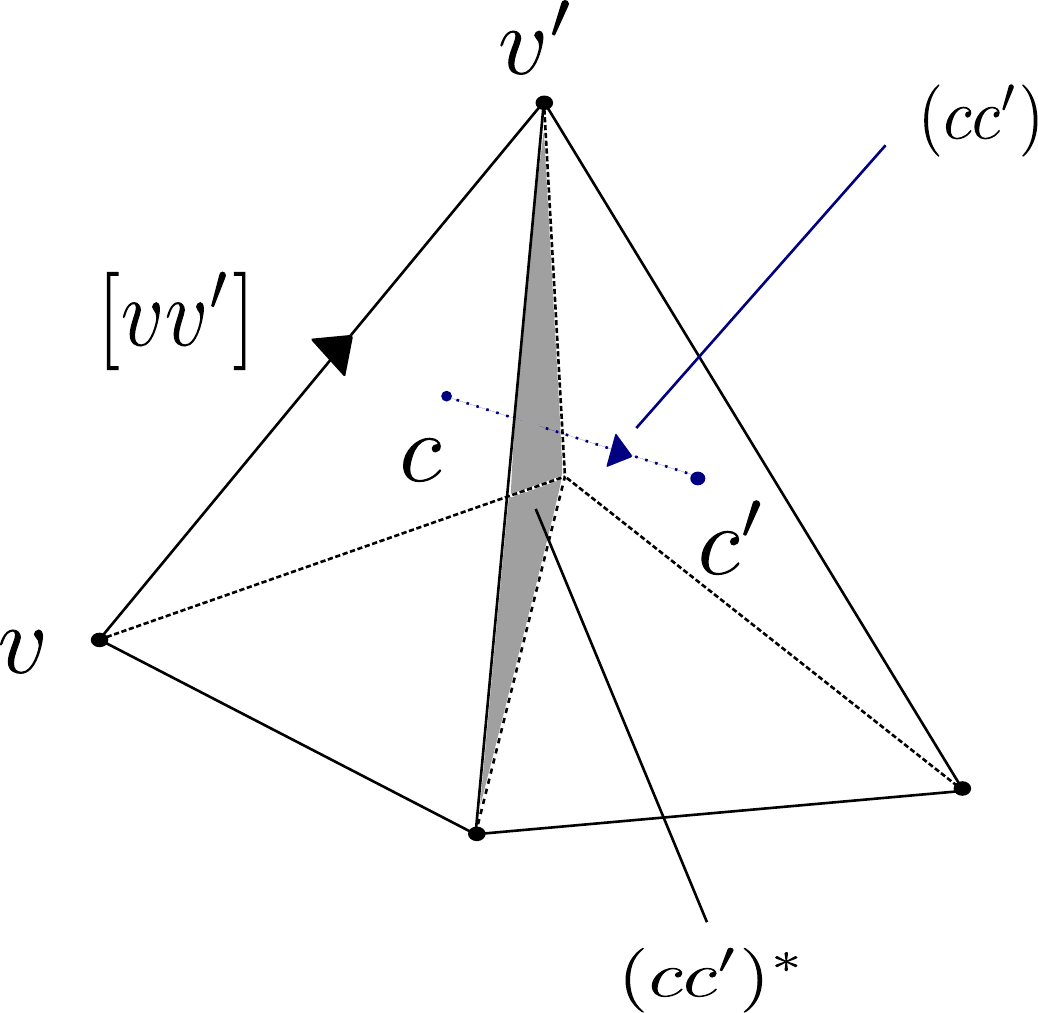}
    \caption{A small piece of the cellular decomposition, showing how we label structures. The centers of the tetrahedron are labelled by $c$ and $c'$ with the connecting link labelled $(cc')$. The triangle which the link passes through is labelled $(cc')^*$. The vertices shown are labelled by $v$ and $v'$ and the connecting edge by $[vv']$. There are arrows on the links and edges indicating the orientation.}
    \label{fig:twotetrahedra}
\end{figure}

 We divide the spatial slice $M$ into subregions forming a cellular decomposition. The 3 dimensional cells will be tetrahedra for simplicity, but other cases can be considered as well. Within each tetrahedron we identify a center point $c$, which we refer to as a node. The tetrahedron dual to $c$ is denoted $c^*$. The oriented segment between two nodes, $c$ and $c'$ is denoted by the ordered pair $(cc')$, which we will call a link. The ordering of the nodes determines the orientation of the link. We refer to the first node in the pair $(cc')$ as the source and the second node as the target. 
 The vertices $\overline{v}$ of tetrahedra are denoted with an overline. The edges between two vertices, $v$ and $v'$ is then denoted $[\overline{vv}']$. For each vertex $v$, there is a set of tetrahedra containing $v$. The centers of these tetrahedra generate a polyhedron in the dual cellular complex. This polyhedron will be denoted $\overline{v}^*$. The faces of the polyhedron, called dual faces are labelled by the edge it intersects, $[\overline{vv}']^*$ or by the set of nodes it contains. Similarly, the triangles making up the surface of the tetrahedra are either labelled by the three vertices they contain such as $[\overline{v}_1\overline{v}_2\overline{v}_3]$ or by the link intercepting it, $(cc')^*$. Some of the structures are shown in figure \ref{fig:twotetrahedra}.

\subsection{Recovering the standard discretization of 4d Euclidean BF theory}\label{sec:BF-disc}
We use the discretization approach that has been used in several works \cite{Freidel:2011ue, Dupuis:2017otn, Dupuis:2020ndx}.

{\textit{\textbf{Restricting the fields to subregions}} }  First, let's consider the symplectic potential of BF theory
\begin{align}
    \Theta_{BF} = \int_M \la \B \wedge \delta\A \ra = \sum_c \int_{c^*} \la \B_c \wedge \delta\A_c \ra
\end{align}
The second equality makes it explicit that we are dividing $M$ into 3-cells $c^*$. Furthermore, we label the restrictions of the fields $\A$ and $\B$  within each cell with a subscript. The fields in neighbouring 3-cells will be related through some continuity relations on the boundary of the cell. 

{\textit{\textbf{Truncation.}} }
Until now we have only rewritten the symplectic potential to account for the fact we broke up $M$ into subregions. The next step is to \textit{truncate} the theory,
by going on shell in the interior of each cell.
Since we are dealing with a topological theory, we could argue that we push all  curvature defects or "torsion" defect (ie such that $d_\A\B \neq0$ to respectively the edges and the vertices of the triangulation. Then we should regularize properly them and treat them accordingly. This would be going beyond the scope of the present paper, so we just assume that there are no such defects at all and leave their study for later investigations.

The equations of motion (\ref{eq: eom}) imply that $\A$ is a flat connection and therefore pure gauge. For an $\ISO(4)$ group element $\cH_c(x)$, interpreted as an ISO(4) holonomy connecting $c$ to a point $x$ in the cell,  
\begin{align}
    \A_c = \cH_c^{-1}d\cH_c
\end{align}
is a solution for $\F = 0$. Then, for a $\mathfrak{iso}(4)^*$ valued 1-form $\chi$,
\begin{align}
    \B_c = \cH_c^{-1}d\chi_c \cH_c
\end{align}
is a solution to $d_\A\B = 0$. 
Hence both $\A_c$ and $\B_c$ are pure gauge.

{\textit{\textbf{Continuity equations.}} } 
As mentioned above, the fields inside each cell $c$ may be considered separately so long as there is continuity between cells. This puts conditions on the fields $\cH$ and $\chi$.

The continuity relations for $\A$ and $\B$ \textit{on the interior of} the triangle shared by $c^*$ and $c'^*$ are expressed as
\begin{align}
    \A_c(x) = \A_{c'}(x), \,  \implies \, \cH_{c'}(x) = \G_{c'c}\cH_{c}(x)\label{cont1}\\
    \B_c(x) = \B_{c'} (x)\implies d\chi_{c'}(x) = \G_{c'c}d\chi_c(x) \G_{cc'} \label{eq: chi cont},
\end{align}
where $\G_{cc'}\in ISO(4)$ does not depend on $x$. We denote the inverse $\G_{cc'}^{-1} = \G_{c'c}$. The above continuity equations are valid on the interior of the triangle. If there is no curvature concentrated on the edges (we will assume this later on), the equations would be valid there as well.

\medskip

We emphasize that we also have the induced equation $[\F,\B]=0$  and  the Bianchi identity $d_\A \F=0$. We should also assess how they can be realized in the truncated scheme.

Since they are expressed in terms of the notion of curvature, they should be obtained by considering several continuity relations concatenated together to generate a loop. To this aim, let us consider the loop $\partial e^*$, which is the boundary of the dual face $e^*$ (dual to the edge $e$). This loop can be described by the links relating the nodes $(c_ic_{i+1})$.  

The  version of the condition $[\F,\B]=0$  in terms of continuity equation is then 
\be
d\chi_{c} = \big(\prod_{(c_ic_{i+1})\in \partial e^*}\G{c_ic_{i+1}}\big)\mone d\chi_c \big(\prod_{(c_ic_{i+1})\in \partial e^*}\G_{c_ic_{i+1}}\big), \label{BF simp constraint}
\ee
As we could infer that the curvature is given in terms of the holonomy around the loop  $\partial e^*$, the Bianchi identity is naturally obtained by demanding that the (dual) polyhedron made of loops $\partial e^*_i$ is closed so that  
\be
\prod_{e_i^*\in\partial v^*} \G_{\partial e_i^*}=1.
\ee
We will not use this constraint in the BF discretization. However we demand  as a requirement for the fields $\chi_{c}$ and $ \G_{c_ic_{i+1}}$ to satisfy \eqref{BF simp constraint}.

\medskip

{\textit{\textbf{Evaluation of the symplectic potential.}} } In order to write the potential in terms of $\cH$ and $\chi$, we should express the variation $\delta \A$ in terms of $\cH$:
\begin{align}
    \delta \A_c = \cH_c^{-1}(d\Delta \cH_c) \cH_c
\end{align}
where $\Delta \cH_c = \delta \cH_c \cH_c^{-1}$.
 The potential evaluated on-shell in the cells  reads
\begin{align}
    \Theta_{BF} \approx \sum_{c} \int_{c^*}\la d\chi_c \wedge d\Delta \cH_c \ra, \label{eq: bf potential}
\end{align}
where $\approx$ means we went on-shell \textit{in} the cells. We can then use the continuity equations \eqref{cont1} and \eqref{eq: chi cont} to simplify its expression and recover the well-known results.

We note that the integrand in \eqref{eq: bf potential}  is a total derivative so we can use Stokes theorem to recast it as an integral over triangles bounding each tetrahedron.  However, there is a choice to be made as to which variable keeps the derivative when dealing with the integral on the boundary. A similar choice arises when dealing with (3d) gravity, and we have the LQG or dual LQG picture \cite{Dupuis:2017otn}. For now we will deal with the case where the derivative is kept on the 1-form $\chi$. 

The standard discretization of the 4d BF theory is summarized in the  following proposition.
\begin{proposition}
The symplectic potential is given as a sum of symplectic potentials associated to the phase space $T^*\ISO(4)$.  
\begin{align}
    \Theta_{BF} = \int_M \la \B \wedge \delta\A \ra \approx  \sum_{(cc')} \la \beta_{(cc')^*}, \Delta \G_c^{c'}\ra,\label{eq: BF potential discrete}
\end{align}
which we construct from the solutions of   the continuity equations \eqref{cont1}  and \eqref{eq: chi cont},
\begin{align}
    \chi_{c'} &= \G_{c'c}(\chi_c +d\cZ_c^{c'})\G_{cc'},  \quad  \cH_{c'}= \G_{c'c}\cH_{c},  
   \text{ and }   \beta_{(cc')^*}= \int_{(cc')^*}d \chi_c. 
\end{align}
 Table  \ref{tab:1} provides the geometric structure  which the discretized fields are attached to.
\begin{table}[h!]
    \centering
    \begin{tabular}{|c|c|c|c|}
    \hline
         Link $(cc')$ & Dual face $e^*$ &  Edges $e$  
         &  Triangles  $(cc')^*$\\
         \hline\hline
         $\G_{cc'}\in\ISO(4)$ &--  &--&
         $\beta_{(cc')^*}\in \iso^*(4)$\\
         \hline
    \end{tabular}
    \caption{Localization of the discrete variables. }
    \label{tab:1}
\end{table}
\\
These discrete variables satisfy by definition two kinds of constraints, the so-called Gauss and ``face simplicity" constraints  
\begin{align}
\sum_{(cc)^*\in\partial c^*}\beta_{(cc)^*}
=0, 
&&
\beta_{(cc)^*}=\big(\prod_{(c_ic_{i+1})\in \partial e^*}\G_{c_ic_{i+1}}\big)\mone \,\beta_{(cc)^*}\, \big(\prod_{(c_ic_{i+1})\in \partial e^*}\G_{c_ic_{i+1}} \big),
\end{align}  
where in the second constraint, the loop being considered begins and ends at the node $c$. Furthermore, if we assume there is no curvature, then we  have the flatness constraint and the discretized Bianchi identity.
 \begin{align}
\G_{e}=\prod_{(c_ic_{i+1})\in \partial e^*}\G_{c_ic_{i+1}}=1 ,   \quad  \prod_{e^*\in\partial v^*}\G_{e}=1.
\end{align}  
\end{proposition}
We note that the flatness constraint implies the face simplicity as well as the discretized Bianchi identity (as it should).

\begin{proof}
Let us evaluate the symplectic potential with the given choice of application of Stokes theorem.  
\begin{align}
    \Theta_{BF} \approx -\sum_{c} \int_{\partial c^*}\la d\chi_c \, ,\, \Delta \cH_c \ra \label{eq: bf potential 1choice}
\end{align}
The boundary of the tetrahedra $c^*$ is made up of four triangles. Since each triangle is shared by two tetrahedra, the contribution to the potential from each triangle contains two terms with a relative minus sign to account for the opposite orientation:
\begin{align}
    \Theta &= \sum_{(cc')} \int_{(cc')^*} \Theta_{(cc')}\\
    \Theta_{(cc')} &\coloneqq \la d\chi_c\, ,\, \Delta \cH_c \ra - \la d\chi_{c'}\, ,\, \Delta \cH_{c'} \ra \label{eq: triangle potential}\\
    &= \la \Delta \G_{c}^{c'} \, ,\, d\chi_c\ra
\end{align}
The last equality is obtained by using the continuity equations and defining $\Delta \G_c^{c'}=\delta \G_{cc'}\G_{c'c}$.  We can identify the factors with structures of the cellular decomposition and its dual graph. We define discrete variables $\G_{(cc')}=\G_{cc'}$ to be the discrete variable associated to the link $(cc')$ and $\beta_{(cc')^*} = \int_{(cc')^*}d\chi_c$ is the discrete variable associated to the triangle $(cc')^*$. Thus, as a function of discrete variables, the potential is
\begin{align}
    \Theta_{BF} \approx \sum_{(cc')} \la \beta_{(cc')^*}, \Delta \G_c^{c'}\ra
\end{align}

From this potential, we can determine the Poisson brackets, which are the canonical ones associated with the cotangent bundle $T^*ISO(4)$. We will review this in the following section.


\textit{{\textbf{ Gauss constraint.}}}
\textit{By construction} the phase space variables satisfy some constraints. For a given tetrahedron, if we perform the sum over the triangles 
\begin{align}
\sum_{(cc_i)^*\in\partial c^*}\beta_{(cc_i)^*}=\sum_{(cc_i)^*\in\partial c^*} \int_{(cc_i)^*} d\chi_c=0  ,   \label{eq: bfconstraint}
\end{align}
by Stokes theorem. This constraint is the discretization of the (pull-back of the) continuum constraint $d_{\cA} \cB=0$. 

In order to accommodate the different possible orientations of the links connecting $c^*$ to its neighbours, we point out that the base point of the variable $\beta$ can be changed according to
\begin{align}
    \beta_{(c'c)^*} = \int_{(c'c)^*} d\chi_{c'} = -\int_{(cc')^*} h_{c'c}d\chi_{c} h_{cc'}= -h_{c'c}\beta_{(cc')^*}h_{cc'}.
\end{align}

\textit{{\textbf{Face simplicity.}}} 
Since we have by the continuity equations that 
\be
d\chi_{c} = \big(\prod_{(c_ic_{i+1})\in \partial e^*}\G_{c_ic_{i+1}}\big)\mone d\chi_c \big(\prod_{(c_ic_{i+1})\in \partial e^*}\G_{c_ic_{i+1}}\big), \ee
where the product of links begins and ends on the node $c$, we can just perform the integration over $(cc')^*$
and get the face simplicity constraint. 
\bes
\beta_{(cc)^*} = \int_{(cc)^*} d\chi_{c} &=\big(\prod_{(c_ic_{i+1})\in \partial e^*}\G_{c_ic_{i+1}}\big)\mone \,\big(\int_{(cc)^*} d\chi_{c}\big) \, \big(\prod_{(c_ic_{i+1})\in \partial e^*}\G_{c_ic_{i+1}} \big) .    
\ees

{\textit{\textbf{Flatness constraint.}}}
The definition of the discretized field does not imply that the holonomies $\G_{c_ic_{i+1}}$ should be flat, so we  implement it by hand. 
\begin{align}
\prod_{(c_ic_{i+1})\in \partial e^*}\G_{c_ic_{i+1}} =1   
\end{align} 
This constraint is the discretization of the (pull-back of the) continuum constraint $\F =0$. One can check that they generate the discretized version of the $BF$ symmetries. We note that this is a non-abelian group valued momentum map \cite{Alekseev3}.

{\textit{\textbf{Bianchi identity.}}}
This condition is naturally discretized by demanding that concatenating all the holonomies on the dual faces of a (dual) polyhedron $v^*$ (as dual to a vertex $v$) gives the identity. This is automatically satisfied if each face is flat. The constraint then reads for every dual polyhedron $v^e$ with faces $e^*$,
 \be
\prod_{e^*\in\partial v^*}\G_{e^*}=1. 
\ee
\end{proof}

\subsection{Relativistic spinning top phase space}\label{sec: spinning top}
We can decompose the discrete variables into subalgebra components. For simplicity we will omit the indices $(cc')$. The $\mathfrak{iso}(4)^*$ holonomy is decomposed as $\beta = b+ V$ for $b \in \so(4)^*$ and $V \in \R^{4*}$. This is equivalent to decomposing the continuous variable as $d\chi_c = db_c+d\sigma_c$ for $b_c \in \so(4)^*$ and $\sigma_c \in \R^{4*}$. We write the $\ISO(4)$ holonomy as $\G = e^{x}h$ where $h\in SO(4)$ and $x$ is a constant element of the Lie algebra of $\R^4$ (which also happens to be $\R^4$). Since $\R^4$ is abelian, in calculations we use the representation such that\footnote{We use the following representation highlighting that the abelian group product of $\R^4$ is isomorphic to the the Lie algebra $\R^4$ seen as an abelian group. The generators $P$ of the Lie algebra $\R^4$ are such that $P^\mu P^\nu=0$. As a consequence the group element is $e^{P^\mu}=1+P^\mu$. We recover in this way the addition as the product of the group $\R^4$, since 
$$
e^{p\cdot P}e^{q\cdot P}= (1+ p\cdot P)(1+q\cdot P)= 1+ p\cdot P+q\cdot P
=1+ (p+q)\cdot P=e^{(p+q)\cdot P}. $$
} $e^{x}=1+x$. The variation then reads $\Delta \G = \delta x+\Delta h+[x,\Delta h]$. Associated to a link $(cc')$,   the potential is
\begin{align}
    \Theta = \la b\, ,\, \Delta h\ra + \la V\, ,\, \delta x\ra+\la V\, ,\, [x,\Delta h]\ra.\label{eq: potential bf}
\end{align}
This is the symplectic potential for the phase space $T^*\ISO(4)$. The Poisson brackets corresponding to this potential are then
\begin{eqnarray}
   & \{ x_\sigma, V^\rho \} = \delta_\sigma^\rho, \quad 
    \{x_\mu, b^{\sigma\rho}\} = 2x_\lambda\eta^{\lambda[\sigma}\delta_\mu^{\rho]}\\
    &\{V^\alpha,b^{\sigma\rho}\} = 2\eta^{\alpha[\rho}V^{\sigma]}, \quad 
    \{h^\alpha{}_\beta, b^{\sigma\rho}\} = (J^{\sigma\rho}h)^\alpha{}_\beta\\
    &\{ b^{\alpha\beta}, b^{\mu\nu} \} = f^{\alpha\beta\mu\nu}{}_{\sigma\rho}(b^{\sigma\rho}+2[x,V]^{\sigma\rho}),
\end{eqnarray}
where $f$ is the structure constant of $\so(4)$: $f^{\alpha\beta\mu\nu}{}_{\sigma\rho} = \demi \la J_{\sigma\rho}^*,[J^{\alpha\beta},J^{\mu\nu}]\ra $.

\begin{figure}
    \centering
    \includegraphics[width=0.3\textwidth]{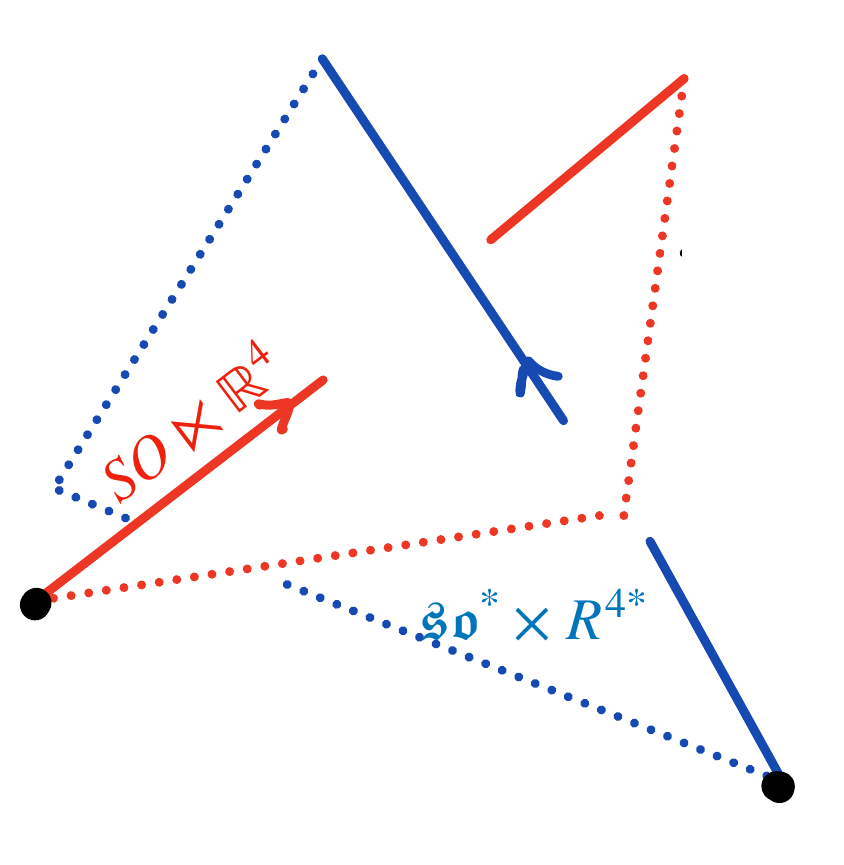}
    \caption{A link in red is decorated by a ISO(4) holonomy, while the triangle in blue is decorated by a $\iso^*(4)\sim \so^*(4)\times \R^{4*}$ element. The building blocks to construct the discrete phase space are given in terms of $T^*\ISO(4)\sim (\SO\ltimes\R^4)\ltimes (\so^*(4)\times \R^{4*})$, with $\so^*(4)\sim \R^6$ and  $\R^{4*}\sim \R^4$. }
    \label{fig:doublewedge1}
\end{figure}

The discrete variables are summarized in table \ref{tab: discrete variables 0} and Fig. \ref{fig:doublewedge1}.
\begin{table}[h!]
    \centering
    \begin{tabular}{|c|c|c|c|}
    \hline
         Link $(cc')$ & Dual face $e^*$ &  Edges $e$  
         &  Triangles  $(cc')^*$\\
         \hline\hline
         $h_{(cc')}\in\SO(4), \, x_{(cc')}\in \R^4$ &  --&--&
         $ b_{(cc')^*}\in \so^*(4), \, V_{(cc')^*}\in \R^4$\\
         \hline
    \end{tabular}
    \caption{Localization of the discrete variables for the BF discretization. }
    \label{tab: discrete variables 0}
\end{table}

Note that the phase space we have recovered is the isomorphic to the phase space of the relativistic spinning top  \cite{Hanson:1974qy}, as one could expect. We define a new variable $S = b + [V,x]$. The potential (for a given link/face pair) now reads
\begin{align}
    \Theta = \la \Delta h \, ,\, S\ra + \la V\, ,\, \delta x\ra \label{eq: top}
\end{align}
The Poisson brackets of these new variables are those of the relativistic spinning top: 
\begin{align}
    \{h{}^\alpha{}_\beta, S^{\mu\nu}\} = (J^{\mu\nu}h){}^\alpha{}_\beta,\quad
    \{S^{\alpha\beta}, S^{\mu\nu}\} = f^{\alpha\beta\mu\nu}{}_{\sigma\rho}S^{\sigma\rho},\quad
    \{x_\mu,V^\nu\} = \delta_\mu^\nu
\end{align}

\medskip

\paragraph{Constraints/charges.} The total phase space we have obtained for the triangulation can therefore be interpreted as a set of relativistic spinning tops, which satisfy some constraints.  
 This is consistent with Penrose's idea of spin networks \cite{Penrose71angularmomentum:}.  We have the constraints encoding the conservation of (relativistic) angular momentum
\begin{eqnarray}
&& \left.\begin{array}{c}  
d_A B +[C\wedge\Sigma]=0\\ d_A\Sigma=0
\end{array}\right\}
\Leftrightarrow d_\A\B=0 \rightarrow \cJ_c=\sum_{(cc)^*\in\partial c^*}\beta_{(cc)^*}=0
\Leftrightarrow 
\left\{ \begin{array}{c}   b_c = \sum_{(cc')\in \partial c^*} b_{(cc')^*} =0\\ 
V_c = \sum_{(cc')\in \partial c^*}V_{(cc')^*} = 0
    \end{array}\right. 
\end{eqnarray} 
The curvature constraints are given by
 \begin{eqnarray}   
  && \left.\begin{array}{c}    
    F=0\\ d_AC=0
\end{array}\right\}
\Leftrightarrow \F=0 \rightarrow \G_{e}=\prod_{(c_i c_{i+1})\in \partial e^*}\G_{c_ic_{i+1}} =1
\Leftrightarrow 
\left\{ \begin{array}{c}   h_{e}= \prod_{(c_i c_{i+1})\in \partial e^*} h_{c_ic_{i+1}} = 1\\ 
x_{e} = \sum_i h_{c_1c_i}x_{(c_ic_{i+1})}h_{c_ic_1} =0
    \end{array}\right.
\end{eqnarray} 
To get the components of $\G_{e}$, we write $\G_{cc'} = e^{x_{(cc')}}h_{cc'}$.  
\begin{align}
 \G_{e}=  \prod_{(c_i c_{i+1})\in \partial e^*} \G_{c_i c_{i+1}} =& \G_{c_1c_2}\dots \G_{c_nc_1}\\
    =&e^{x_{(c_1c_2)}}h_{c_1c_2}e^{x_{(c_2c_3)}}h_{c_2c_3}\dots e^{x_{c_nc_1}}h_{c_nc_1} \\
    =& e^{x_{(c_1c_2)}}h_{c_1c_2}e^{x_{(c_2c_3)}}h_{c_2c_1}h_{c_1c_2}h_{c_2c_3}\dots e^{x_{c_nc_1}}h_{c_nc_1}\\
    =& e^{\sum_i h_{c_1c_i}x_{(c_ic_{i+1})}h_{c_ic_1}}\prod_{(c_i c_{i+1})\in \partial e^*} h_{c_ic_{i+1}} = e^{x_{e}} h_{e}.
\end{align}
where $n$ in the above is the total number of nodes around the edge $e$ and we define $c_{n+1} = c_1$.

Finally we also have the Bianchi identity for $F$ and its companion. The  Bianchi identity $d_\A\F=0$ implies the constraints
 \begin{eqnarray}   
  && \left.\begin{array}{c}    
    d_AF=0\\ d_A(d_AC)=[F\wedge C]=0
\end{array}\right\}
\rightarrow \prod_{e^*\in\partial v^*}\G_{e}=1
\Leftrightarrow 
\left\{ \begin{array}{c}  \prod_{e^*\in\partial v^*} h_{e^*} = 1\\ 
 \sum_{e^*\in\partial v^*} x_{e^*}=0
    \end{array}\right.
\end{eqnarray} 
The face simplicity constraints follows from the constraint $d_\A\B =0$:
 \begin{eqnarray}   
   \left.\begin{array}{c}    
    [F\wedge B]+[d_\A C\wedge\Sigma]=0\\ d_A(d_A\Sigma)=[F\wedge\Sigma]=0
\end{array}\right\}
\Leftrightarrow d_\A (d_\A\B)=0 \rightarrow \beta_{(cc)^*}=\G_{e}^{-1} \,\beta_{(cc)^*}\G_{e}    
\Leftrightarrow 
\left\{ \begin{array}{l}  
b_{(cc)^*}=h_{e}^{-1} (b_{(cc)^*} + [V_{(cc)^*},x_e]) h_{e} \\ 
V_{(cc)^*}= h_{e}^{-1} \,V_{(cc)^*}h_{e}
    \end{array}\right.\nonumber
\end{eqnarray} 

\section{Discretization of BFCG theory}\label{sec: discbfcg}

In this section we will go through the same procedure starting from the BFCG potential. To illustrate that the theory is similar to the ISO(4) BF theory, we will first show how to recover the results of the previous subsection (the discrete potential (\ref{eq: potential bf})). We will then proceed and   obtain the proper discretized BFCG potential, the main result of this section. As a consequence we will recover the classical picture behind the G-networks introduced in \cite{Asante:2019lki}.

\paragraph{Restricting the fields to subregions. }
We start by expressing the integral $\Theta_{BFCG}$ as a sum of integrals over each cell.
\begin{align}
    \Theta_{BFCG} = \sum_{c} \int_{c^*}\la B_c\wedge \delta A_c  \ra -\sum_c\int_{c^*} \la C_c\wedge \delta \Sigma_c \ra. 
\end{align}

\paragraph{Truncation. } As before, we go on-shell inside the cells. 
Since BFCG and BF theory differ by a boundary term, they share the same equations of motion so we can either decompose the solutions of (\ref{eq: eom}) or directly solve (\ref{eq: bfcgeom}). We will take the former approach: we solve the equations of motion from BF theory to get, for a $\ISO(4)$ holonomy $\cH_c(x)$ connecting $c$ to $x$ in the cell and a $\mathfrak{iso}(4)^*$ valued 1-form $\chi$,
\begin{align}
    \A_c = \cH_c^{-1}d\cH_c, \quad 
    \B_c = \cH_c^{-1}d\chi_c \cH_c.
\end{align}
We decompose $\cH=e^cg$, where $g$ is a rotation and $e^c$ is a translation. $\chi$ is decomposed into  $\chi = b + \sigma$ for $b \in \so(4)^*$ and $\sigma \in \R^{4*}$. Thus, (still using the convenient representation $e^c=1+c$)
\begin{align}
    \A_c=A_c+C_c = g_c^{-1}dg_c +g_c^{-1}dc_c g_c && \B_c=B_c+\Sigma_c = g_c^{-1}(db_c+[d\sigma_c,c_c]+d\sigma_c)g_c
\end{align}
giving
\begin{align}
    A_c =& g^{-1}_cdg_c & C_c =& g_c^{-1}dc_c g_c\\
    \Sigma_c =& g_c^{-1}d\sigma_c g_c & B_c =& g_c^{-1}(db_c +[d\sigma_c , c_c])g_c,
\end{align}
where the different fields are defined in table \ref{tab: cont eq}. 

\paragraph{Continuity equations. }
The continuity of the field between neighbouring cells $c^*$ and $c'^*$ is expressed as
\begin{align}
    B_c = B_{c'} && A_c = A_{c'} && C_c = C_{c'} && \Sigma_c = \Sigma_{c'} &&\text{on }c\cap c'^*\label{eq: cont1}
\end{align}
The solutions of these continuity equations are given in Table \ref{tab: cont eq}. 

If we apply the continuity equations consecutively around a loop $\partial e^*$, we also get the equations
\begin{align}  \label{cont eq loop}
& dc_c = h_{e}\mone dc_c h_{e},   \\ 
&  db_{c} =h_{e}\mone db_{c}h_{e}, \quad  d\sigma_{c} =h_{e}\mone d\sigma_{c}h_{e}, \quad h_e\equiv \prod_{c_ic_{i+1}\in\partial e^*}h_{c_ic_{i+1}},\label{cont eq loop1}
\end{align} 
where again, the product along the loop of links begins and ends at node $c$. The condition \eqref{cont eq loop} can be seen as the discretization of   $[F,C]=0$, while the second ones \eqref{cont eq loop1} come from $d_\A\B=0$.  The solutions of the continuity equations are given in the table \ref{tab: cont eq}. 
\begin{table}[h!]
    \centering
    \begin{tabular}{|c|c|c|}
    \hline
         Continuity eqs & Solutions to continuity eqs  &  Fields 
         \\
         \hline\hline
$g_c^{-1}dg_c = g^{-1}_{c'}dg_{c'}$&$  g_c = h_{cc'}g_{c'}$ & \makecell{ $g_c$ function in $\SO(4)$,\\ $h_{cc'}$ a constant  in $\SO(4)$  }     \\
\hline
$      dc_{c'} = h_{c'c}dc_c h_{cc'}$        & $ c_{c'} = h_{c'c}(c_c+x_c^{c'})h_{cc'}$  &\makecell{$c_c$     function  in $\R^{4}$,\\  $x_c^{c'}$  constant in $\R^4$ } \\
\hline
  $d\sigma_{c'} = h_{c'c}d\sigma_c h_{cc'}$ & $\sigma_{c'} = h_{c'c}(\sigma_c + d\varsigma_c^{c'})h_{cc'}$ &       \makecell{$\sigma_c$      1-form in $\R^{4*}$,\\  $\varsigma_c^{c'}$ function in $\R^{4*}$}\\
  \hline
    $db_{c'} =h_{c'c}(db_c -[d\sigma_c,x_c^{c'}])h_{cc'}$ & $b_{c'} =h_{c'c}(b_c -[\sigma_c , x_c^{c'}]+dy^{c'}_c)h_{cc'} $ & \makecell{$b_c$     1-form in $\so(4)^*$,\\  $y_c^{c'}$ function in $\so(4)^*$}\\
         \hline
    \end{tabular}
    \caption{Continuity equations and their solution. }
    \label{tab: cont eq}
\end{table}
Anticipating a bit, we will see that in the BFCG discretization, if we allow for some curvature on the edges,  we will still need  to assume that on the edges of the triangulation 
\begin{align}\label{sigmasimple0}
    \sigma_c= h_e^{-1} \sigma_c h_e. 
\end{align}
This will ensure that we can integrate the symplectic potential. While in the construction given in \cite{Asante:2019lki} particular emphasis was put on the \textit{edge simplicity}, that is something of the type 
\begin{align}
    c_c= h_e^{-1} c_c h_e, 
\end{align}
to recover the KBF amplitude, 
it seems that for the discretization, the key property will be more \eqref{sigmasimple0}.

\paragraph{Evaluating the symplectic potential: before making the choice.}
In order to express the potential in terms of the fields $g, \sigma, b, $ and $c$ we need to express the variation of the fields $A$ and $\Sigma$ in these variables. We define
\begin{align}
    \Delta g_c \coloneqq \delta g_c g_c^{-1} 
\end{align}
then,
\begin{align}
    \delta A_c = g^{-1}_cd\Delta g_c g _c && \delta \Sigma_c = g^{-1}(\delta d\sigma_c + [d\sigma_c,\Delta g_c])g_c
\end{align}
Using these expressions, the potential in a cell is
\begin{align}
    \Theta_c \approx  \la db_c\wedge d\Delta g_c \ra + d\la [d\sigma_c, c_c], \Delta g_c \ra - \la dc_c\wedge d\delta \sigma_c \ra, \label{eq: potential c}
\end{align}
where $\approx$ means we went on-shell. We see that $\Theta_c$ is a total derivative and can be written as an integral over the boundary $\partial c^*$ by Stokes theorem. As in the previous section, there is a choice to make regarding which variable keeps the derivative when we perform Stokes' theorem.

\subsection{Choice 1: recovering the   BF discretization}
At this time, we make the following choice (we note that the first term is the same polarization as the LQG case \cite{Dupuis:2017otn}). 
\begin{align}
    \Theta_{BFCG} \approx \sum_{c} \int_{\partial c*} (\la db_c \, ,\, \Delta g_c \ra +\la [d\sigma_c , c_c] \, ,\, \Delta g_c \ra -\la c_c\, ,\, d \delta \sigma_c \ra)
\end{align}
The boundary $\partial c^*$ is made up of four triangles. Each triangle is shared by two tetrahedra. 
The contribution from each triangle, $(cc')^*$,  is $\Theta _{(cc')^*}$,
\begin{align}
    \Theta_{BFCG}  &\approx \sum_{(cc')^*} \int_{(cc')^*}\Theta _{(cc')^*}\\
    \Theta_{(cc')^*} &=\la db_c \, ,\, \Delta g_c \ra - \la db_{c'}\, ,\, \Delta g_{c'}\ra  + \la [d\sigma_c,c_c]\, ,\, \Delta g_c\ra - \la [d\sigma_{c'},c_{c'}]\, ,\, \Delta g_{c'}\ra \nonumber\\
                     &\qquad - \la c_c\, ,\, d\delta \sigma_c\ra + \la c_{c'} \, ,\, d\delta \sigma_{c'} \ra.
\end{align}

\begin{proposition}
The symplectic potential is given as a sum of symplectic potential associated to the phase space $T^*\ISO(4)$.  
\begin{align}
    \Theta_{BFCG} \approx   \Theta'_{BF}  = \sum_{(cc')^*} \left\la \Delta h_c^{c'}\, ,\, \int_{(cc')^*} db_c \right\ra + \left\la [\Delta h_c^{c'} , x_c^{c'}] \, ,\, \int_{(cc')^*}d\sigma_c \right\ra + \left\la x_c^{c'}\, ,\, \delta \int_{(cc')^*} d\sigma_c \right\ra , 
\end{align}
where the discrete variables are obtained from the continuity equations from table \ref{tab: cont eq},

 Table  \ref{tab:3} provides the geometric structure  which they are attached to.
\begin{table}[h!]
    \centering
    \begin{tabular}{|c|c|c|c|}
    \hline
         Links & Dual faces&  Edges  &  Triangles  \\
         \hline\hline
          $h_{(cc')}\in\SO(4)$,  $x_{(cc')}\in\R^4$   &-- &-- & $b_{(cc')^*}=\int_{(cc')^*}db_c\in\so^*(4)$,   $V_{(cc')^*}=\int_{(cc')^*}d\sigma_c$
         \\
         \hline
    \end{tabular}
    \caption{Localization of the discrete variables. }
    \label{tab:3}
\end{table}
\end{proposition}
We note that we almost recover the same potential as in the BF standard discretization \eqref{eq: potential bf}. The difference comes from a minus sign in the $(x,V)$ sector. The reason is the following. By adding the boundary term to go to the BFCG action, we have swapped the polarization, we have exchanged the configuration and momentum variables. If we consider a symplectic form $\delta q \wedge \delta p$, this form is not invariant under the exchange $q\leftrightarrow p$, which leads to $\delta p \wedge \delta q=-\delta q \wedge \delta p$. This is what we have done by changing the polarization. The symplectic transformation is instead given  by $q\dr -p$ and  $p\dr q$. Hence the $V$ in the BF discretization and the $V$ in the BFCG discretization are related by a minus sign.

\begin{proof}
We can use the continuity relations from table \ref{tab: cont eq} to simplify $\Theta_{(cc')^*}$.
\begin{align}
    \Theta_{(cc')^*} = \la db_c \, ,\, \Delta h_c^{c'} \ra +\la x_c^{c'} \, ,\, [d\sigma_c , \Delta h_c^{c'}]\ra +\la x_c^{c'}\, ,\, \delta d\sigma_c \ra,
\end{align}
where we have defined $\Delta h_c^{c'} \coloneqq \delta h_{cc'} h_{c'c}$. 

The total potential is now
\begin{align}
    \Theta_{BFCG}  &\approx \sum_{(cc')^*} \left\la \Delta h_c^{c'}\, ,\, \int_{(cc')^*} db_c \right\ra + \left\la [\Delta h_c^{c'} , x_c^{c'}] \, ,\, \int_{(cc')^*}d\sigma_c \right\ra + \left\la x_c^{c'}\, ,\, \delta \int_{(cc')^*} d\sigma_c \right\ra\\
    &\approx \sum_{(cc')}\la \Delta h_{(cc')}\, ,\, b_{(cc')^*}\ra +\la [\Delta h_{(cc')}, x_{(cc')}]\, ,\, V_{(cc')^*}\ra +\la x_{(cc')}\, ,\, \delta V_{(cc')^*}\ra.\label{eq: potential bfbfcg}
\end{align}
The factors in $\Theta_{BFCG} $ can be associated to structures in the cellular decomposition. We already saw that $h_{cc'}$ is related to the the links in the dual cellular decomposition. Similarly $x_c^{c'}$ is also associated to the links. The factors involving integrals over triangles are associated to the triangles in the cellular decomposition, which are dual to the links. The discrete variables and where they live in the cellular decomposition is summarized in table \ref{tab:3} which is the same as table \ref{tab:1}. 

\end{proof}

\subsection{Choice 2: recovering the phase space of \cite{Asante:2019lki}}
As we emphasized already, when using Stokes' theorem, there is a choice which is made on which variable will keep the differential. 
We recall that we obtained the symplectic potential for a given tetrahedron $c^*$,
\begin{align}
    \Theta_c =  \la db_c\wedge d\Delta g_c \ra + d\la [d\sigma_c, c_c]\, ,\, \Delta g_c \ra - \la dc_c\wedge d\delta \sigma_c \ra .
\end{align}
In the previous section we used Stokes' theorem to write the potential on the triangles bounding $c^*$. In particular, the last term was expressed as
\begin{align}
    \la dc_c \wedge d\delta \sigma_c\ra = d\la c_c \, ,\, d\delta \sigma_c \ra
\end{align}
and when we determined the discrete variables, $\int_{(cc')^*}d\sigma$ was assigned to a triangular face. We can alternatively write this term as 
\begin{align}
    \la dc_c \wedge d\delta \sigma_c \ra = -d\la dc_c\wedge \delta \sigma_c \ra
\end{align}
In performing Stokes' theorem in this way, we will have  different discrete variables living on different structures in the cellular decomposition. 
Let us first identify the discretized variables and then the constraints
associated to them. 
\medskip

\subsubsection{Identifying the discrete variables.} After performing Stokes theorem and applying the continuity equations the potential on the triangles is now 
\begin{align}
    \int_{(cc')^*}\Theta_{(cc')^*} = \int_{(cc')^*}\la d(b_c +[c_c,\sigma_c])\, ,\, \Delta h_c^{c'}\ra -\int_{(cc')^*}d\la (dc_c \, ,\,  [\Delta h_c^{c'},\varsigma_c^{c'}])\ra +\int_{(cc')^*}d\la dc_c \, ,\, \delta \varsigma_c^{c'}\ra\label{eq: potential triangle}
\end{align}
The last term is indeed a problematic one. In contrast to the previous section, neither $dc_c$ nor $\delta\varsigma_c^{c'}$ are constant and so we must do some work to perform this integration. We shall once again use Stokes' theorem on each triangle and deal with integrals over edges bounding triangles instead.
\begin{align}
    \sum_{(cc')^*}\int_{(cc')^*}d\la dc_c\, ,\, \delta \varsigma_c^{c'}\ra=&\sum_{(cc')}\int_{\partial (cc')^*}\la dc_c\, ,\,  \delta \varsigma_c^{c'}\ra \\
    =&\sum_e \int_e  \sum_{(cc')\in e^*}\epsilon^e_{(cc')}\la d c_c\, ,\, \delta \varsigma_c^{c'}\ra 
\end{align}
The first sum and the integral are over edges $e$. The second sum is over the links $(cc')$ which make up the polygon $e^*$ dual to the $e$. The factor $\epsilon^e_{(cc')}$ is either 1 or $-1$, depending on whether the orientation of $ (cc')^*$ is aligned with $e$ or not. 

\begin{figure}
    \centering
    \includegraphics[width=0.2\textwidth]{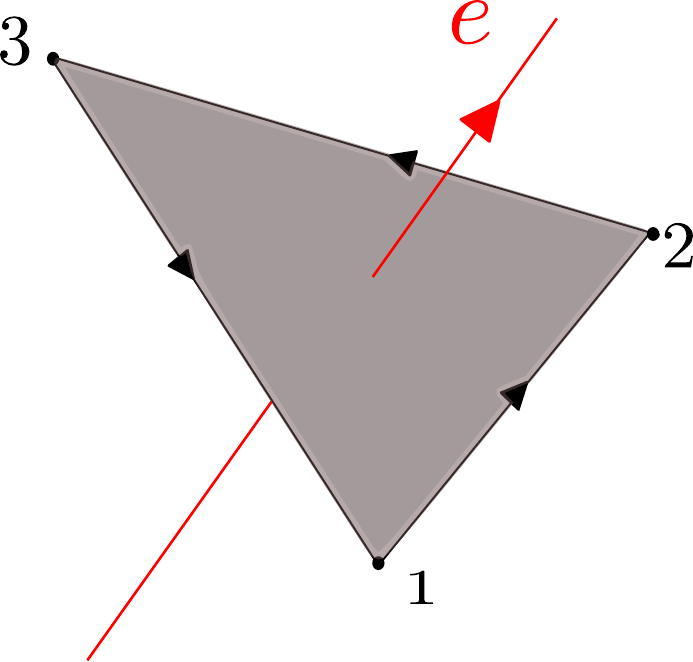}
    \caption{An example of the type of edge used for illustrative calculations. The edge is shown in red labelled by $e$ with surrounding nodes forming a triangle. }
    \label{fig:edge}
\end{figure}

To illustrate we  take the example edge we have in figure \ref{fig:edge}. The contribution of this edge to the potential is
\begin{align}
    \int_e \sum_{(cc')^*\in e^*} \la dc_c \, ,\, \delta \varsigma_c^{c'} \ra =& \int_e \la dc_1\, ,\, \delta \varsigma_1^2 \ra +\int_e\la dc_2 \, ,\, \delta\varsigma_2^3\ra + \int_e \la dc_3\, ,\, \delta \varsigma_3^1 \ra  \\
    =& \int_e \la dc_1 \, ,\, (\delta \varsigma_1^2 +h_{12}\delta \varsigma_2^3 h_{21} + h_{13}\delta\varsigma_3^1 h_{31}) \ra\\
    =& \int_e \la dc_1 \, ,\, \delta (\varsigma_1^2 +h_{12}\varsigma_2^3 h_{21} +h_{13}\varsigma_3^1h_{31}) \ra+\int_e \la [dc_1 , h_{12}\varsigma_2^3 h_{21}]\, ,\, \Delta h_1^2\ra \nonumber\\
    &+\int_e \la [dc_1,h_{13}\varsigma_3^1 h_{31}]\, ,\, \Delta h_1^3\ra \label{eq: edge potential}
\end{align}
In the second line, we were able to use the continuity equation in the variable $c$ to base each term at the center 1 (an arbitrary choice). 
The second and third term involve something proportional to $\Delta h_1^2$ and $\Delta h_1^3$ (the value of the superscript and subscript are a result of the arbitrary choice made to base everything at 1) and can therefore be absorbed in the first term of (\ref{eq: potential triangle}) (since the total potential involves summing over the links). This will be the source of a non-trivial closure constraint for the tetrahedron. 

The first term involves a combination of the continuity variables $\varsigma$.

Here, as we alluded earlier, we need to take a specific assumption on the behaviour of $\sigma_c$ under consecutive change of frames, as in \eqref{sigmasimple0}. 

Consider the three continuity equations for $\sigma$ which are satisfied on $e = (12)^*\cap (23)^*\cap (31)^*$:
\begin{align}
    \sigma_2 = h_{21}(\sigma_1 +d\varsigma_1^2)h_{21} && \sigma_3 = h_{32}(\sigma_2+d\varsigma_2^3)h_{23}&& \sigma_1 = h_{13}(\sigma_3+d\varsigma_3^1)h_{31}.
\end{align}
Putting these together we have
\begin{align}
    \sigma_3 =& h_{32}h_{21}h_{13}\sigma_3 h_{31}h_{12}h_{23} + h_{32}h_{21}h_{13}d\varsigma_3^1 h_{31}h_{12}h_{23} \nonumber\\&+  h_{32}h_{21}d\varsigma_1^2 h_{21}h_{23} +h_{32}d\varsigma_2^3h_{23}.\label{eq: edge continuity}
\end{align}
\textit{Assuming}\footnote{If we enforce the fact there is \textit{no} curvature on the edges, this assumption is obviously true. If, we do not impose flatness right away, we have to make this assumption to get to the relevant result. Hence we can get our result, even though there is no flatness, but such that our assumption is satisfied. } that 
\begin{align}\label{eq: edge continuity1}
    \sigma_3 =& h_{32}h_{21}h_{13}\sigma_3 h_{31}h_{12}h_{23}.
\end{align}
and putting together (\ref{eq: edge continuity}) and \eqref{eq: edge continuity1}, we get
\begin{align}
    d\varsigma_1^2+h_{13}d\varsigma_3^1h_{31} + h_{12}d\varsigma_2^3 h_{21} = 0.
\end{align}
And so we have that 
\begin{align}
     \varsigma_1^2+h_{13}\varsigma_3^1h_{31} + h_{12}\varsigma_2^3 h_{21} = V^{e^*}_1,
\end{align}
for some constant $V_1^{e^*}$, which then decorates the dual face $e^*$. This is exactly the expression which appears in the first term of (\ref{eq: edge potential}). Since $V^{e^*}$ is a constant, we are able to consider $\int_e dc_1$ as our discrete variable associated to $e$ and $V^{e^*}_1$ as the discrete variable associated to the polygon $e^*$. The potential due to $e$ is then
\begin{align}
    \int_e \sum_{(cc')^* \in e^*} \la dc_c \, ,\, \delta \varsigma_c^{c'} \ra =& \la \delta V_1^{e^*} \, ,\, \int_e dc_1 \ra +\int_e \la [dc_1 , h_{12}\varsigma_2^3 h_{21}]\, ,\, \Delta h_1^2\ra  + \int_e \la [dc_1,h_{13}\varsigma_3^1 h_{31}]\, ,\, \Delta h_1^3\ra .
\end{align}
Summarizing,  the symplectic potential takes now the shape
\begin{align}
    \Theta_{BFCG} \approx \sum_{(cc')} \left\la \int_{(cc')^*} \tilde b_{c}^{c'}\, ,\, \Delta h_c^{c'} \right\ra+\sum_e \left\la \int_e dc_{c_e}\, ,\, \delta V^{e^*}_{c_e}\right\ra\label{eq: potentialbfcg}
\end{align}
A lot has been concealed in writing equation (\ref{eq: potentialbfcg}). The label $c_e$ is the choice of base point in the polygon dual to $e$ (in the example edge we took $c_e$ to be the node 1). We also introduced $\tilde b_c^{c'}$. Simply put, this is shorthand notation for everything which appears in $\Theta$ next to $\Delta h_c^{c'}$. The explicit form of such a term depends on the choices of $c_e$ and so we won't write it out in general. We will define $\tilde b_{c}^{c'}$ in an explicit example shortly. 

\begin{figure}
    \centering
    \includegraphics[width=0.3\textwidth]{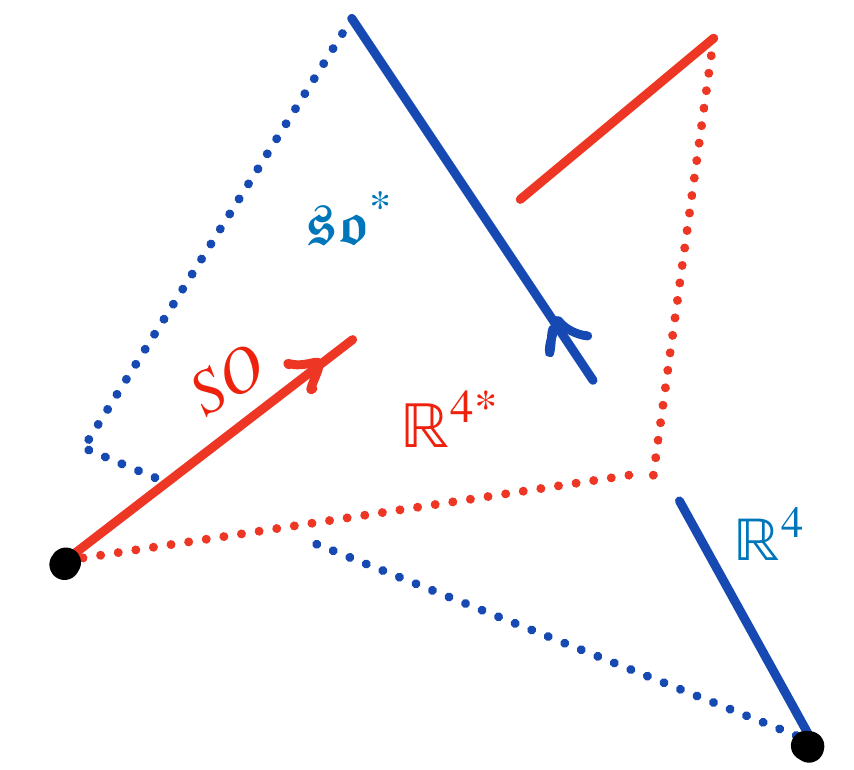}
    \caption{A link in red is decorated by a SO(4) holonomy, a dual face is decorated by an element in $\R^{4*}\cong \R^4$. An edge in blue is decorated by an element in $\R^4$,  the triangle in blue is decorated by a $\so^*(4)\cong \R^{6}$ element. The building blocks to construct the discrete phase space are still isomorphic to $T^*\ISO(4)$ since $(\SO(4)\ltimes \R^{4*})\ltimes (\so^*\times \R^4)\cong (\SO(4)\ltimes \R^{4})\ltimes (\so^*\times \R^{4*})\cong T^*\ISO(4)$. }
    \label{fig:doublewedge2}
\end{figure}

We can now determine the discrete variables. The discrete variables are $b_{(cc')^*}=\int_{(cc')^*} \tilde b_c^{c'} \in \so(4)^*$ on triangles, $h_{cc'}\in \SO(4)$ on links dual to triangles, $\ell_e = \int_e dc_{c_e} \in \R^4$  on edges, and $V_{e^*} = V_{c_e}^{e^*}$ on polygons dual to an edge. The variables are summarized in table \ref{tab: bfcgl} and Fig. \ref{fig:doublewedge2}. 
\begin{table}[h!]
    \centering
    \begin{tabular}{|c|c|c|}
    \hline
         Discrete variable & Definition in terms of & Home in cellular complex \\
         &continuous variables & \\\hline\hline
         $V_{c_e}$ & linear combination of $\varsigma$'s  & Polygon $e^*$, a face in \\
         & around an edge & the dual complex\\ \hline
         $\ell_{c_e}$ & $\int_e dc_{c_e}$ & Edge $e$ of tetrahedron\\ \hline
         $h_{(cc')}$ & $h_{cc'}$ & Links\\\hline
         $b_{(cc')^*}$ & $\int_{(cc')^*} 
         \tilde b_c^{c'}$ & Triangles \\\hline
    \end{tabular}
    \caption{Summary of discretization of BFCG theory. The key result is that $b_{(cc')^*}$ depends on many variables, namely $c, \varsigma$ and $b$, as illustrated in \eqref{bs}. In particular, we have integrations both on the triangle and some of the edges forming its boundary. }
    \label{tab: bfcgl}
\end{table}

The discrete potential looks just like that of (\ref{eq: top}), but with the translation sector on the edges and dual faces instead of the links and triangles. 
\begin{align}
    \{\ell_\alpha, V^\beta \} &= -\delta _\alpha^\beta\label{poi1}\\
    \{h^\alpha{}_\beta ,b^{\sigma\rho}\} &= (J^{\sigma\rho}h)^\alpha{}_\beta\quad 
    \{b^{\sigma\rho},b^{\alpha\beta}\} = \eta^{\sigma\alpha}b^{\rho\beta}+\eta^{\rho\beta}b^{\sigma\alpha}-\eta^{\sigma\beta}b^{\rho\alpha}-\eta^{\rho\alpha}b^{\sigma\beta}\label{poi2}
\end{align}
For each edge $e$ of the triangulation we have the decorations $\ell_e$ and $V_{e^*}$ which decorate respectively the edge and the dual face, with Poisson bracket \eqref{poi1}. For each link $l$, we have the decorations $h_l$ and $b_l$ which decorate respectively the link  and the triangle $l^*$, with Poisson brackets \eqref{poi2}.

These variables satisfy a  set of constraints different than in the ones in the BF discretization. In order to justify these constraints and see that they follow from the definitions of the discrete variables in terms of the continuous functions, we need to explicitly define $\tilde b_c^{c'}$ and make choices about where to base the edge variables. In order to simplify these expressions and be exhaustive, we will consider an explicit example.

\subsubsection{Explicit case: example of the sphere}
\begin{figure}
    \centering
    \includegraphics[width=0.4\textwidth]{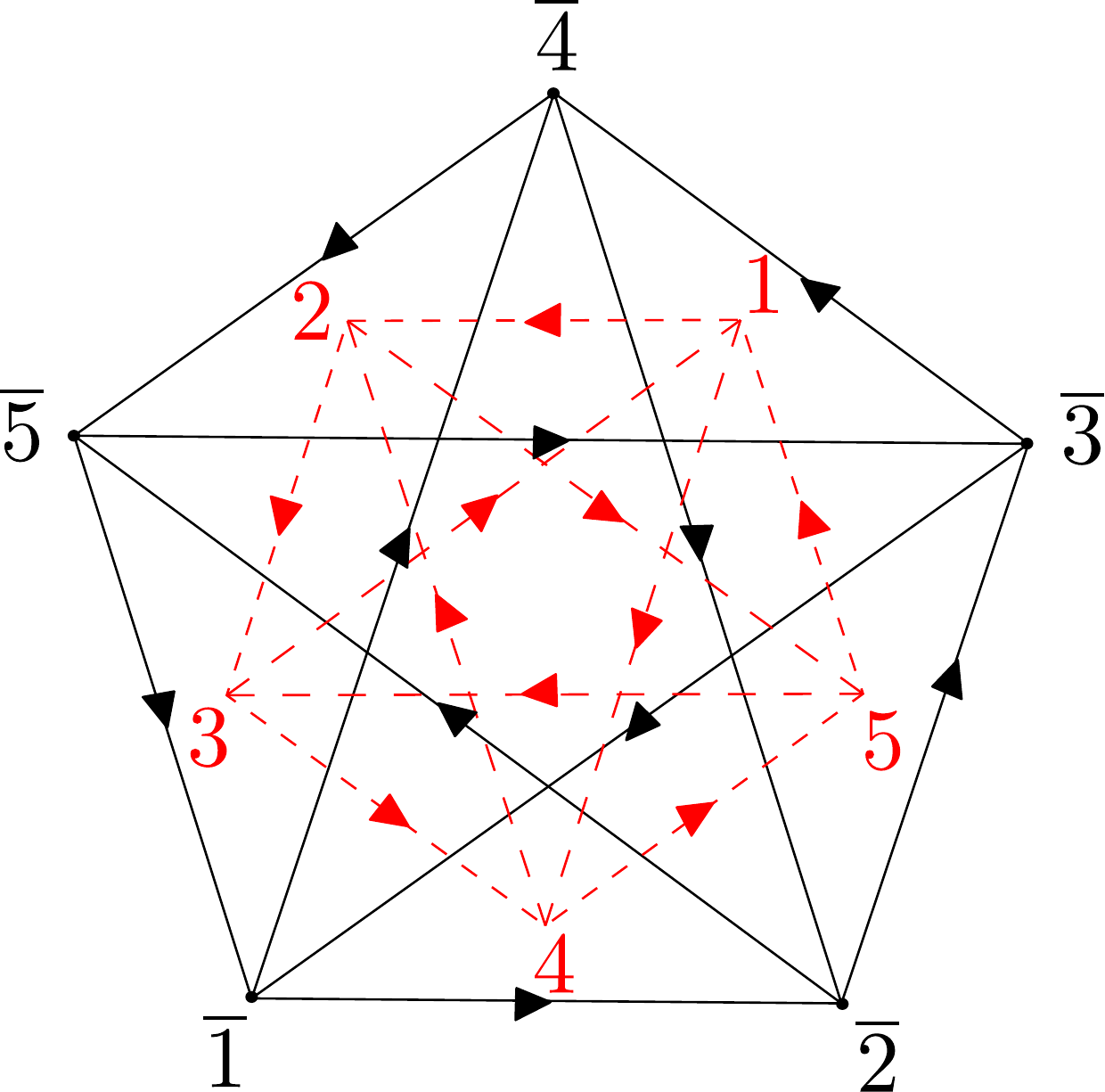}
    \caption{
    The edges of the complex are shown with solid black lines and the dual complex is shown with dotted red lines. The arrows indicate the orientation chosen.}
    \label{fig:4cel}
\end{figure}

The triangulation we choose is that of the 4-simplex. The space is divided into 5 tetrahedra. The centers of the 5 tetrahedra will be labelled by integers $\{1, 2,\dots, 5\}$. The vertices of the tetrahedra will be labelled by overlined integers $\{\overline{1},\overline{2},\dots,\overline{5}\}$. The tetrahedron $i^*$ will have vertices $\{\overline{1},\dots, \overline{5}\}\backslash \overline{i}$. A diagram indicating the orientation of the links and edges is shown in figure \ref{fig:4cel}. The calculation on the edges is just like in the example we did previously since each edge is dual to a triangle.  The resulting dual face variables are 
\begin{align}
         V_3^{[\overline{12}]^*} = \varsigma_3^4+h_{34}\varsigma_4^5+h_{35}\varsigma_5^3 &&  V_2^{[\overline{31}]^*} = -h_{24}\varsigma_4^2+h_{24}\varsigma_4^5-\varsigma_2^5\nonumber\\
         V_2^{[\overline{14}]^*} = \varsigma_2^3-\varsigma_2^5-h_{25}\varsigma_5^3&& V_2^{[\overline{51}]^*} = h_{24}\varsigma_4^2+h_{23}\varsigma_3^4+\varsigma_2^3\nonumber\\
         V_1^{[\overline{23}]^*} = h_{14}\varsigma_4^5+h_{15}\varsigma_5^1+\varsigma_1^4&&
         V_3^{[\overline{42}]^*} = -h_{35}\varsigma_5^3-\varsigma_3^1+h_{35}\varsigma_5^1\nonumber\\
         V_1^{[\overline{25}]^*} = h_{13}\varsigma_3^4-h_{13}\varsigma_3^1-\varsigma_1^4&&
         V_1^{[\overline{34}]^*} = \varsigma_1^2+h_{15}\varsigma_5^1+h_{12}\varsigma_2^5\nonumber\\
         V_1^{[\overline{53}]^*} = -h_{14}\varsigma_4^2-\varsigma_1^4+\varsigma_1^2&&
         V_1^{[\overline{45}]^*} = \varsigma_1^2+h_{12}\varsigma_2^3+h_{13}\varsigma_3^1\label{Vs}
\end{align}
In the above, we have always based the variables at the lowest node in numerical order. This is a choice made arbitrarily, any node which is a vertex of $[\overline{ij}]^*$ would be equally valid. The resulting expressions for the $\so(4)^*$ variables are:
\begin{align}
     b_{(12)^*} =& \int_{[\overline{345}]} (db_1-d[c_1,\sigma_1]-[dc_1,d\varsigma_1^2]) +\int_{[\overline{45}]}[dc_1,h_{12}\varsigma_2^3h_{21}]+\int_{[\overline{34}]} [dc_1,h_{12}\varsigma_2^5h_{21}]\nonumber\\
    b_{(23)^*} =& \int_{[\overline{145}]} (db_2-d[c_2,\sigma_2]-[dc_2,d\varsigma_2^3]) +\int_{[\overline{51}]}[dc_2,h_{23}\varsigma_3^4h_{32}]\nonumber\\
    b_{(34)^*} =& \int_{[\overline{125}]} (db_3-d[c_3,\sigma_3]-[dc_3,d\varsigma_3^4]) +\int_{[\overline{12}]} [dc_3,h_{34}\varsigma_4^5h_{43}]\nonumber\\
    b_{(45)^*} =& \int_{[\overline{123}]} (db_4-d[c_4,\sigma_4]-[dc_4,d\varsigma_4^5]) \nonumber\\
    b_{(51)^*} =& \int_{[\overline{234}]} (db_5-d[c_4,\sigma_4]-[dc_5,d\varsigma_5^1]) -\int_{[\overline{34}]}[dc_5,\varsigma_5^1]-\int_{[\overline{23}]}[dc_5,\varsigma_5^1]\nonumber\\
    b_{(14)^*} =& \int_{[\overline{235}]} (db_1-d[c_1,\sigma_1]-[dc_1,d\varsigma_1^4])+\int_{[\overline{23}]}[dc_1,h_{14}\varsigma_4^5h_{41}]-\int_{[\overline{53}]} [dc_1,h_{14}\varsigma_4^2h_{41}]\nonumber\\
    b_{(25)^*} =& \int_{[\overline{134}]} (db_2-d[c_2,\sigma_2]-[dc_2,d\varsigma_2^5])-\int_{[\overline{14}]}[dc_2,h_{25}\varsigma_5^3h_{52}]\nonumber\\
    b_{(31)^*} =& \int_{[\overline{245}]} (db_3-d[c_3,\sigma_3]-[dc_3,d\varsigma_3^1])+\int_{[\overline{25}]}[dc_3,\varsigma_3^1]-\int_{[\overline{25}]}[dc_3,\varsigma_3^4]-\int_{[\overline{45}]}[dc_3,\varsigma_3^1]\nonumber\\
    b_{(42)^*} =& \int_{[\overline{135}]}(db_4-d[c_4,\sigma_4]-[dc_4,d\varsigma_4^2])-\int_{[\overline{51}]}[dc_4,\varsigma_4^2]-\int_{[\overline{31}]} [dc_4,\varsigma_4^5] +\int_{[\overline{31}]}[dc_4,\varsigma_4^2]\nonumber\\
    b_{(53)^*} =& \int_{[\overline{124}]}(db_5-d[c_5,\sigma_5]-[dc_5,d\varsigma_5^3])-\int_{[\overline{42}]}[dc_5,\varsigma_5^1]+\int_{[\overline{42}]}[dc_5,\varsigma_5^3]-\int_{[\overline{12}]} [dc_5,\varsigma_5^3] \label{bs}
\end{align}
Clearly there is a lack of symmetry due to the orientation choices for the links.

\subsubsection{Constraints}
The discrete variables in this new polarisation give rise to a new set of constraints. We follow the terminology of \cite{Asante:2019lki} to name the constraints.

Compared to the BF case, there are two sets of new constraints, the 2-Gauss constraints encoding the triangles are closed and the 2-flatness encoding that the dual polyhedra close.

We then have the more usual set of constraints, the 1-Gauss constraints and the 1-flatness constraints. The latter encodes that the holonomies along the links forming a closed loops should be trivial. The former encodes that the triangle decoration should be equal to a specific quantity.

Finally, there is also the edge simplicity constraints. This constraint is actually implied if we have flatness, but can hold without having flatness. 

\medskip 

Let us review the explicit shape of the constraints. 

\medskip 

\paragraph{2-Gauss constraints.}

For each triangle in the simplex, there is a constraint on the edge data. For example, for the  triangle $(45)^*$ 
\begin{align}
    h_{43}\ell_3^{[\overline{12}]}h_{43} + h_{41}\ell_1^{[\overline{23}]}h_{14}+ h_{42}\ell_2^{[\overline{31}]}h_{24} = \int_{\partial [\overline{123}]} dc_4 = 0.
\end{align}
Such constraint is the discrete analogue of the constraint $d_A C =0$, since  
\begin{align}
C=g\mone dc  g \Leftrightarrow dc = g C g\mone \Leftrightarrow d^2c=0=g(d C + [g\mone d g,C ])g\mone =    g(d C + [A,C ])g\mone.
\end{align}

In general, the edge variables $\ell$ corresponding to the edges of a triangle sum to zero. The sum can only be performed after each variable is transported to the appropriate node. The full list of triangle constraints in the 4-simplex are given below
\begin{align}
    \G^{[\overline{123}]}_4 =& h_{ 43 } \ell_3 ^{ [ \overline{12} ] } + h_{41} \ell_1^{ [ \overline{ 23 }] } + h_{ 42 } \ell_2^{ [\overline{31}] }\nonumber\\
    \G^{[\overline{124}]}_5 =& h_{ 53 } \ell_3 ^{ [ \overline{12} ] } + h_{51} \ell_1^{ [ \overline{ 25 }] } + h_{ 52 } \ell_2^{ [\overline{51}] }\nonumber\\
    \G^{[\overline{125}]}_3 =& \ell_3 ^{ [ \overline{12} ] } - \ell_3^{ [ \overline{ 42 }] } - h_{ 32 } \ell_2^{ [\overline{14}] }\nonumber\\
    \G^{[\overline{134}]}_2 =& - \ell_2 ^{ [ \overline{31} ] } -  \ell_2^{ [ \overline{ 14 }] } + h_{ 21 } \ell_1^{ [\overline{34}] }\nonumber\\
    \G^{[\overline{135}]}_4 =& -h_{ 42 } \ell_2 ^{ [ \overline{31} ] } - h_{41} \ell_1^{ [ \overline{ 53 }] } + h_{ 42 } \ell_2^{ [\overline{51}] }\nonumber\\
    \G^{[\overline{145}]}_2 =&\ell_2 ^{ [ \overline{14} ] } + h_{21} \ell_1^{ [ \overline{ 45 }] } + \ell_2^{ [\overline{51}] }\nonumber\\
    \G^{[\overline{234}]}_5 =& h_{ 51 } \ell_1 ^{ [ \overline{23} ] } + h_{53} \ell_3^{ [ \overline{ 42 }] } + h_{ 51 } \ell_1^{ [\overline{34}] }\nonumber\\
    \G^{[\overline{235}]}_1 =&  \ell_1 ^{ [ \overline{23} ] } - \ell_1^{ [ \overline{ 53 }] } +  \ell_1^{ [\overline{25}] }\nonumber\\
    \G^{[\overline{245}]}_3 =& -h_{ 31 } \ell_1 ^{ [ \overline{25} ] } -  \ell_3^{ [ \overline{ 42 }] } + h_{ 31 } \ell_1^{ [\overline{45}] }\nonumber\\
    \G^{[\overline{345}]}_1 =& \ell_1 ^{ [ \overline{34} ] } +  \ell_1^{ [ \overline{ 45 }] } +  \ell_1^{ [\overline{53}] }\label{Gs}
\end{align}
The relative sign differences between terms comes from the $\epsilon^e_{(cc')}$ factor introduced above.

\medskip 

\paragraph{2-flatness.}
This set of constraints is interpreted as the closure of the dual polyhedra (tetrahedra in our illustrating example).  By Minkowski's theorem, the sum of vectors normal to faces of a polyhedron with magnitude equal to the face area is zero. In this system, the analogous quantities are the variables on the faces, $V$. For example the closure of the polyhedron dual to $\overline{1}$ is
\begin{align}
    h_{23}V^{[\overline{12}]^*}_3h_{32} +V_2^{[\overline{14}]^*}+V_2^{[\overline{13}]^*}+V_2^{[\overline{15}]^*} =0
\end{align}
This follows directly from the definitions of the face variables (also using $V^{[\overline{ij}]}=-V^{[\overline{ji}]}$). In our example, there are five such polyhedron constraints shown below
\begin{align}
     \cP^{\overline{1}} =& h_{23}V^{[\overline{12}]^*}_3h_{32} +V_2^{[\overline{14}]^*}-V_2^{[\overline{31}]^*}-V_2^{[\overline{51}]^*} \nonumber\\
     \cP^{\overline{2}} =& V^{[\overline{23}]^*}_1 +V_1^{[\overline{25}]^*}-h_{13}V_3^{[\overline{12}]^*}h_{31}-h_{13}V_3^{[\overline{24}]^*}h_{31}\nonumber \\
     \cP^{\overline{3}} =& h_{13}V^{[\overline{31}]^*}_3h_{31} +V_1^{[\overline{34}]^*}-V_2^{[\overline{53}]^*}-V_2^{[\overline{23}]^*}\nonumber\\
     \cP^{\overline{4}} =& V^{[\overline{45}]^*}_1 +V_1^{[\overline{42}]^*}-h_{12}V_2^{[\overline{14}]^*}h_{21}-V_1^{[\overline{34}]^*}\nonumber\\
     \cP^{\overline{5}} =& V^{[\overline{53}]^*} +h_{12}V_2^{[\overline{51}]^*}h_{21}-V_1^{[\overline{45}]^*}-V_1^{[\overline{25}]^*} \label{Ps}
\end{align}

\medskip 

\paragraph{1-Gauss.}
Due to the less trivial continuity equations for the $\so(4)^*$ variables, the expression for the tetrahedron constraints appears more cumbersome:
\begin{align}
    \sum_{c':c\to c'}b_{(cc')^*}-\sum_{e}[\ell_c^{e}, V_c^{e^*}] - \sum_{c': c'\to c} h_{cc'}b_{ (c'c)^* } h_{c'c} = 0 
\end{align}
The explicit form of the sum depends on where we choose to base our variables $\ell$ and $V$ as well as the orientation of the links. The first sum is a sum over the links which have the source point at $c$ and the third sum is over links which have their target node at $c$. The second sum is over the edges such that $\ell^e$ and $V^{e^*}$ is based at $c$. If no $\ell$ is based at $c$, this is just the empty sum. With the variables defined above, we have the five constraints
\begin{align}
   \cT^{5} =&  b_{ (51)^* } + b_{ (53)^* } -h_{54} b_{ (45)^* } h_{45} -h_{52}b_{ (35)^* } h_{25}\nonumber \\
    \cT^{4} =& b_{ (45)^* } + b_{ (42)^* } -h_{41} b_{ (14)^* } h_{41} -h_{43}b_{ (34)^* } h_{34}\nonumber\\
    \cT^{3} =& b_{ (34)^* } + b_{ (31)^* } -h_{32} b_{ (23)^* } h_{32} -h_{35}b_{ (53)^* } h_{53} - [\ell^{ [\overline{12}]}_3, V^{[\overline{12}]^*}_3 ] - [\ell_3^{[\overline{42}]}, V_3^{ [\overline{42}]^*} ]\nonumber\\
    \cT^2  =&b_{ (23)^* } + b_{ (25)^* } -h_{24} b_{ (42)^* } h_{42} -h_{21}b_{ (12)^* } h_{21} - [\ell^{ [\overline{14}]}_2, V^{[\overline{14}]^*}_2 ] - [\ell_2^{[\overline{31}]}, V_2^{ [\overline{31}]^*} ] - [\ell_2^{[\overline{51}]}, V_2^{ [\overline{51}]^*} ]\nonumber\\
    \cT^1 =& b_{(12)^*} + b_{(14)^*} -h_{13}b_{(31)^*}h_{31} -h_{15}b_{(51)^*}h_{51} -[\ell_1^{ [ \overline{45} ] } , V_1^{ [ \overline{45} ]^* }]-[\ell_1^{ [ \overline{23} ] } , V_1^{ [ \overline{23} ]^* }]\nonumber\\
    &-[\ell_1^{ [ \overline{53} ] } , V_1^{ [ \overline{53} ]^* }]-[\ell_1^{ [ \overline{34} ] } , V_1^{ [ \overline{34} ]^* }] -[\ell_1^{ [ \overline{25} ] } , V_1^{ [ \overline{25} ]^* }]\label{Ts}
\end{align}
We emphasize that these constraints are realized by definition of the fields.
We expect them to be the discretization of the constraint $d_AB+[C\wedge \Sigma]=0$.
\medskip

Since these constraints do not seem to be very natural, let us illustrate in one example how this comes to be. For concreteness, let's take $\cT^3$ as an example. For convenience we recall the relevant triangle variables 
\begin{align}
    b_{(23)^*} =& \int_{[\overline{145}]} (db_2-d[c_2,\sigma_2]-[dc_2,d\varsigma_2^3]) +\int_{\overline{51}}[dc_2,h_{23}\varsigma_3^4h_{32}]\\
    b_{(34)^*} =& \int_{[\overline{125}]} (db_3-d[c_3,\sigma_3]-[dc_3,d\varsigma_3^4]) +\int_{[\overline{12}]} [dc_3,h_{34}\varsigma_4^5h_{43}]\\
     b_{(31)^*} =& \int_{[\overline{245}]} (db_3-d[c_3,\sigma_3]-[dc_3,d\varsigma_3^1])+\int_{\overline{25}}[dc_3,\varsigma_3^1]-\int_{\overline{25}}[dc_3,\varsigma_3^4]-\int_{\overline{45}}[dc_3,\varsigma_3^1]\\
     b_{(53)^*} =& \int_{\overline{124}}(db_5-d[c_5,\sigma_5]-[dc_5,d\varsigma_5^3])-\int_{\overline{42}}[dc_5,\varsigma_5^1]+\int_{\overline{42}}[dc_5,\varsigma_5^3]-\int_{[\overline{12}]} [dc_5,\varsigma_5^3]
\end{align}
Using the continuity equations, we can check that
\begin{align}
    -h_{32} b_{(23)^*} h_{23} = \int_{[\overline{145}]}(-db_3+d[c_3,\sigma_3])-\int_{[\overline{51}]} [dc_3,\varsigma_3^4]
\end{align}
and
\begin{align}
    -h_{35}b_{(53)^*}h_{53} = \int_{[\overline{124}]} (-db_3 +d[c_3,\sigma_3])+\int_{[\overline{42}]} [dc_3,h_{35}\varsigma_5^1h_{53}]-\int_{[\overline{42}]}[dc_3,h_{35}\varsigma_5^3]+\int_{[\overline{12}]} [dc_3,h_{35}\varsigma_5^3h_{53}].
\end{align}
We now evaluate the sum involving the $b$ variables, 
\begin{align}
    b_{ (34)^* } + b_{ (31)^* }& -h_{32} b_{ (23)^* } h_{32} -h_{35}b_{ (53)^* } h_{53}= \int_{[\overline{125}]}d(b_3-[c_3,\sigma_3]) + \int_{[\overline{245}]}d(b_3-[c_3,\sigma_3])\nonumber\\
    &-\int_{[\overline{145}]}d(b_3-[c_3,\sigma_3])-\int_{[\overline{124}]}d(b_3-[c_3,\sigma_3])-\int_{[\overline{125}]}[dc_3,d\varsigma_3^4]
    -\int_{[\overline{245}]}[dc_3,\varsigma_3^1]\nonumber\\
    &+\int_{[\overline{12}]}[dc_3, h_{34}\varsigma_4^5 h_{43}]
    +\int_{\overline{25}}[dc_3,\varsigma_3^1]-\int_{\overline{25}}[dc_3,\varsigma_3^4]-\int_{\overline{45}}[dc_3,\varsigma_3^1]\nonumber\\
    &+\int_{[\overline{42}]} [dc_3,h_{35}\varsigma_5^1h_{53}]-\int_{[\overline{42}]}[dc_3,h_{35}\varsigma_5^3]+\int_{[\overline{12}]} [dc_3,h_{35}\varsigma_5^3h_{53}]-\int_{[\overline{51}]} [dc_3,\varsigma_3^4]\\
    &= \int_{[\overline{1245}]} d^2(b_3-[c_3,\sigma_3])+ \int_{\partial[\overline{125}]}[dc_3,\varsigma_3^4]+\int_{\partial[\overline{245}]}[dc_3,\varsigma_3^1]\nonumber\\
    &+\int_{[\overline{12}]}[dc_3, h_{34}\varsigma_4^5 h_{43}]
    +\int_{\overline{25}}[dc_3,\varsigma_3^1]-\int_{\overline{25}}[dc_3,\varsigma_3^4]-\int_{\overline{45}}[dc_3,\varsigma_3^1]\nonumber\\
    &+\int_{[\overline{42}]} [dc_3,h_{35}\varsigma_5^1h_{53}]-\int_{[\overline{42}]}[dc_3,h_{35}\varsigma_5^3]+\int_{[\overline{12}]} [dc_3,h_{35}\varsigma_5^3h_{53}]-\int_{[\overline{51}]} [dc_3,\varsigma_3^4].
\end{align}
We have used the observation that the four surfaces we are integrating over in the first four terms of the first line are the boundary of the tetrahedron $1^*$. We then use Stokes theorem to write it as derivative in the bulk of the tetrahedron. We can then use that $d^2=0$ to say that the first integral after the second equality vanishes. Next we write out the integral $\int_{\partial[\overline{125}]}$ and $\int_{\partial[\overline{245}]}$ as integrals over edges. We then collect terms:
\begin{align}
     b_{ (34)^* } + b_{ (31)^* } -h_{32} b_{ (23)^* } h_{32} -h_{35}b_{ (53)^* } h_{53} =& \int_{[\overline{12}]} [dc_3, \varsigma_3^4 +h_{34}\varsigma_4^5h_{43}+h_{35}\varsigma_5^3h_{53}]\nonumber\\
     &+ \int_{[\overline{42}]}[dc_3,  h_{35}\varsigma_5^1 h_{53}-h_{35}\varsigma_5^3-\varsigma_3^1]\\
     =& [\ell_3^{[\overline{12}]},V_3^{[\overline{12}]^*}] +  [\ell_3^{[\overline{42}]},V_3^{[\overline{42}]^*}],
\end{align}
which is precisely $\cT^3=0$.

\paragraph{Edge simplicity and 1-flatness}
Now let's consider the edge simplicity constraint. We consider this condition instead of the usual flatness constraint because it appears as the less strict condition we must impose in order to discretize the potential.

The edge simplicity constraint for a given edge $e$ follows directly from the property of the fields we introduced in \eqref{cont eq loop}. By integrating this expression on the edge, we just get
\begin{align}
    \cE_c^e = \ell_c^e -h_e^{-1} \ell_c^eh_e=0, \quad  h_{e}= \prod_{(c_i c_{i+1})\in \partial e^*} h_{c_ic_{i+1}} .\label{edge}
\end{align}
This constraint can be viewed as the discretization of the constraint $[F\wedge C]=0$. 

Similar constraints hold when $V$ replaces $\ell$, which is inherited from \eqref{eq: edge continuity1}. This can be seen as a dual face simplicity.

The 1-flatness constraint appears by constraining the constants $h_{cc'}$ to form a flat closed holonomy along the loop $\partial e^*$. 
\begin{align}
 h_{e}= \prod_{(c_i c_{i+1})\in \partial e^*} h_{c_ic_{i+1}} = 1\end{align}
This constraint implies the constraint \eqref{edge}.

\section{2-group structures}\label{sec: 2gp}
In this section, we put our findings in light of the higher gauge theory context and its discretization. We first recall quickly the concept of higher gauge theory mostly relying on \cite{Baez:2010ya}.  

\subsection{From 2-group to higher gauge theory in a nutshell}
There is a natural 2-group interpretation for the discretized symmetries we have obtained. We recall that a (strict) 2-group can be seen as a crossed module, $(G,H, t, \rhd)$, which  consists in a pair of (Lie) groups $(G,H)$, with a group homomorphism $t:H\dr G$ called the target map and an action $\rhd$ of $G$ on $H$. The target map and the action must satisfy some compatibly relations.
\bes\label{comp cond}
&& t(h)\rhd h'=hh'h\mone, \quad t(g\rhd h)= gt(h)g\mone.  
\ees
The crossed module is equipped with \textit{two} product laws. The first one is inherited from considering $H >\!\!\! \lhd G $.
\be
(h_1\ , \ g_1)\bullet (h_2\ , \ g_2)= (h_1 (g_1\rhd h_2)\ ,\ g_1g_2), \quad (h\ , \ g)^{-1_\bullet} =  ( (g\rhd h)\mone \ , \ g\mone ).
\ee 
The other multiplication comes from the product of $H$.
\bes
&&(h_1\ , \ g_1)\diamond (h_2\ , \ g_2)= (h_1h_2, g_1), \quad \textrm{if } g_2 = t(h_1)g_1. \\
&& (h,g)^{-1_\diamond} = (h\mone, t(h)g), \textrm{ and unit } (1,g).
\ees
We note that demanding that the $t$ map is trivial, the vertical composition implies that the holonomy $g_2g_1\mone$ must be flat. 

\medskip

The notion of 2-Lie algebra which can be exponentiated to the Lie 2-group \cite{Baez:2003fs} is relevant to discuss the \textit{BF} or \textit{BFCG} theories. 

A Lie 2-algebra is given by the differentiation of the Lie 2-group \cite{Baez:2003fs}. A Lie 2-algebra can be given by the differential crossed module  $(Lie\, G, Lie\, H, \tau, \alpha)$, where both $\tau$ and $\alpha$ are obtained by differentiating  $t$ and $\rhd$ respectively. The compatibility relations are now 
\begin{align}
    \tau (\alpha(x)(y))= [x,\tau(y)], \quad \alpha(t(y))(y')=[y,y'], \quad x\in Lie\, G, \quad y,y'\in Lie\, H.  
\end{align}

\medskip 

Equipped with the notion of Lie 2-group and Lie 2-algebra, we can define the notion of higher gauge theory \cite{baez:2004in}. It is specified in terms of a 1-connection, the usual gauge connection, denoted $A$, a 1-form with value in $Lie\, G$, and a 2-connection, often noted $B$ or $\Sigma$, a 2-form with value in $Lie\, H$. Together with   these connections, we get the associated curvatures. We have the 1-curvature, which is the usual curvature found in gauge theory, $F(A)=d A + \frac{1}{2}[A\wedge A]$. We also have the 2-curvature which is given by $G(\Sigma,A)= d \Sigma + \alpha(A)(\Sigma)$, where we used the action of  $Lie\, G$ on $Lie\, H$. We note that the 1-curvature is actually is specified in terms of $\tau(\Sigma)$, 
\begin{align}
    F=\tau(\Sigma).
\end{align}
Finally, we have the 1-gauge transformations and the 2-gauge transformations, parameterized  by a group element $g\in G$ and a 1-form $a\in Lie H$ respectively. 
\begin{align}\label{2-gaugetr}
A'&= g^{-1} A g + g^{-1}d g + \tau(a), \\
\Sigma'&= \alpha(g)(\Sigma)+ d_Aa +a \wedge a, 
\end{align}

\subsection{$BF$ case.} 
The four dimensional  $BF$ action can related to higher gauge symmetries, more specifically to the (co-)tangent 2-group, and at the continuum level to the associated  (co-)tangent 2-Lie algebra \cite{Baez:2010ya}. 

\subsubsection{Continuum level and 2-Lie algebra}
The $\cB$-field, a 2-form with value in $\g^*\sim \R^d$ is usually interpreted as a Lagrange multiplier implementing the fact we are dealing with a flat connection. In the higher gauge theory picture,  we interpret it instead   as a  2-connection, while     $\cA$  with value   in $\g$ is the 1-connection. The relevant 2-Lie group is given by the (co-)tangent 2-group, with $ H=Lie\, G^*\sim \R^d$, $G$, with $G$ acting with the coadjoint action on  $Lie\, G^*$ and the t-map is constant, $t=1$. The connections are valued in the associated Lie 2-algebras. 

The translational symmetry of the $BF$ action is then interpreted as the 2-gauge transformation.

\subsubsection{Discrete level and 2-Lie group}
While the relevant 2-group is (co-)tangent 2-group, we have obtained at the discrete level that this 2-group can actually be seen as a pair of trivial 2-groups, ie a pair of 1-groups. 

Firstly on the dual complex,  we have solely  a decoration of the links, by the group elements in $G$. There are no decoration on the dual faces. This means that the  2-group is actually trivial, we have\footnote{By $H=1$, we mean that the group is simply the identity element. While the $t\equiv1$, we mean that to any element in $H$, the $t$ map associates the identity element in $G$. } 
\begin{align}
    G\equiv ISO(4), \quad H= 1, \quad t\equiv 1, \quad \rhd \textrm{ trivial}
\end{align}

We note that the target map being trivial is also equivalent to the dynamical constraint that the $\ISO(4)$ holonomy on the links must be flat. 

\medskip 

Secondly, on the triangulation, we have no decoration on the edges (so we can set equivalently any decoration to the identity), but the triangles decorated by the abelian group $\R^{10}\sim \iso(4)^*$. \begin{align}
    G\equiv 1, \quad H= \R^{10}\sim \iso(4)^*, \quad t\equiv 1, \quad \rhd \textrm{ trivial}.
\end{align}  

\medskip

Finally, the two (trivial) 2-groups can be seen as dual to each other, one being configuration, the other one momentum space, and together form the (co-)tangent 2-group. Put together, we can set up a phase space for each (link, trivial face decoration)/(trivial edge decoration, triangle), with symplectic form given by \eqref{eq: potential bf}.  

Let us reformulate the previous statement. Locally the cotangent bundle $T^*G\sim \R^d \rtimes G$ has a simple cross module structure (with a trivial t-map) hence can be interpreted as a 2-group. Since the   cotangent bundle $T^*G$ is naturally equipped with a symplectic form, the 2-group is also equipped with a symplectic form. This points towards a possible generalization of the Heisenberg double \cite{SemenovTianShansky:1993ws} to the 2-group context.   

\subsection{$BFCG$ case }

\subsubsection{Continuum level and 2-Lie algebra}
While the \textit{BFCG} action we considered is actually equivalent to a \textit{BF} action, the 2-gauge structure is actually different. Now the 2-connection is not given by the full $\cB=B+\Sigma$ field with value in $Lie\, G\sim \iso(4)^*$, but by the translational sector $\Sigma$ only. The $B$-field component with value in $\so(4)^*$ is now seen  as a Lagrange multiplier.  The 1-gauge connection is not the full connection $\cA=A+ C$ but is $A$ lying in the  $\so(4)$ sector while the $C$ component is seen as a Lagrange multiplier.  As a consequence this means that the relevant gauge symmetry is the Euclidean 2-group $\SO(4)\ltimes \R^4$. 

While this might look as a cosmetic change with respect to the original Euclidean BF theory, it is actually deeper than that. Indeed, the addition of the boundary term modifying the Euclidean BF theory into the BFCG theory can be seen as dualization, more exactly semi-dualization, since we only dualize ``half" of the Lie algebra $\iso(4)$, namely the translational part. While the charge algebra is not modified, the place where these charges will be discretized is modified. 

Said otherwise, in the \textit{BF} formulation, the $\cB$-field, the 2-connection, is seen solely as the momentum, and the 1-connection is solely configuration variable.  In contrast,  in the \textit{BFCG} case, we have that the momentum variables are a mix of 1-connection $C$ (in $\R^4$) and 2-form $B$ (in $\so(4)^*$), while the configuration variables are given in terms of the  1-connection $A$ (in $\so(4)$) and the 2-connection $\Sigma$ in $\R^4$. 

\medskip 

Nevertheless due to the equivalence with the Euclidean BF theory, we still have the (co)-tangent 2-group structure present. To see how they still connect together we can move to the discrete picture.

\subsubsection{Discrete level and 2-Lie groups}

Firstly, on the dual complex, we have $SO(4)$ holonomies decorating on the links and elements in $\R^4$ decorating the dual faces. We recognize this as the Euclidean 2-group. 
\begin{align}
    G\equiv SO(4), \quad H= \R^4, \quad t\equiv 1, \quad \rhd \textrm{ canonical action of } SO(4) \textrm{  on } \R^4. 
\end{align}
Once again the target map being trivial is also equivalent to the dynamical constraint stating that the $SO(4)$ holonomy on the links must be flat.

\medskip 

Secondly, on the triangulation, we have the edges are decorated by elements in $\R^{4*}\sim \R^{4}$, and the triangles decorated by the abelian group $\R^{6}\sim \so(4)^*$. Hence we have the trivial cross module
\begin{align}
    G\equiv \R^{4}, \quad H= \R^{6}\sim \so(4)^*, \quad t\equiv 1, \quad \rhd \textrm{ trivial}.
\end{align}  

\medskip

Finally, once again, the two 2-groups can be seen as dual to each other. Put together, we can set up a phase space for each (link, face )/(edge, triangle), with symplectic form given by \eqref{eq: potentialbfcg}.  The total phase space is again based on $T^*\ISO(4)$. This is where the equivalence with the usual Euclidean $BF$ formulation appears.   
As we recalled in the previous section $T^*\ISO(4)$ can be seen as a cross module equipped with a symplectic structure. Unlike in the \textit{BF} case, now the  components of this 2-group are themselves non-trivial 2-groups in the sense that both contain decoration on the face/triangle. This puts further arguments that the generalization of a 2-Heisenberg double, defined as a 2-group seen as a phase space with both configuration and momentum spaces being 2-groups might exist.

\begin{figure}
    \centering
    \includegraphics[width=0.8\textwidth]{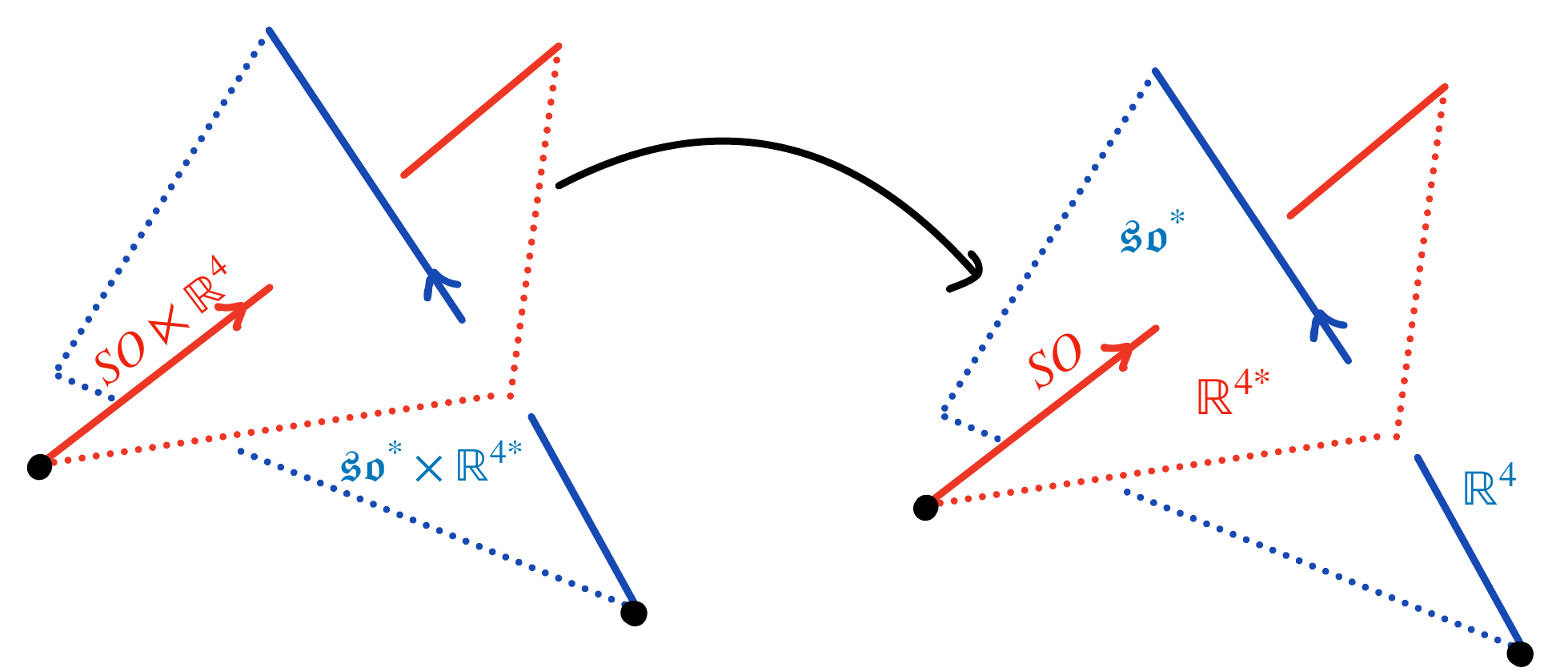}
    \caption{Pictorial representation of the semi-dualization process relating the BF discretization to the BFCG one. }
    \label{fig:doublewedge3}
\end{figure}

\section{Outlook}
We have revisited the discretization of  4d $BF$ theory and highlighted that this discretization (and in turns the quantum theory) is very sensitive on the choice of boundary terms. This is yet another example of importance such boundary data to probe the quantum regime of a gauge theory \cite{Freidel:2020xyx, Freidel:2020svx, Freidel:2020ayo}.  Here the boundary term just implemented a partial change of polarization which allowed to rewrite the Euclidean $BF$ theory as the $BFCG$ theory. It was already known that a full change of polarization leads at the discretized level to the notion of ``$BF$ vacuum" \cite{Dittrich:2014wda}, so that the kinematical symmetries are not the Lorentz ones (through the Gauss constraint) but instead the translational ones (through the flatness constraint).   

Here we performed only a partial change of polarization, in the case where the gauge group is not the Lorentz group but the Euclidean group. At the level of the action, this implemented a change from a regular $BF$ action to the $BFCG$ action. At the discrete level, one gets the usual phase space built out of $T^*\ISO(4)$ for each link of the dual 1-complex, which is the classical phase space behind the standard notion of spin networks. With the partial change of polarization, we obtain the phase space of the so-called $G$-networks, which are based on 2-group structures. We were therefore able to recover the expression  of the $G$-network discrete variables in terms of the continuum variables.

\medskip

Such identification is actually very important to relate gravity to the \textit{BF} or \textit{BFCG} theory. Indeed, gravity can be obtained at the continuum level by constraining the $B$ field, through the simplicity constraints \cite{Mikovic:2011si, Mikovic:2015hza, Belov:2018uko}. 

In the \textit{BFCG} case, what stands for the discretized \textit{B} field is actually a function of several of the continuous fields. Hence demanding that such discretized \textit{B} field  satisfies the simplicity constraint is not equivalent to the continuum simplicity constraint. This might explain why the discrete simplicity constraint leads to such drastic reduction of the model \cite{Mikovic:2011si, Mikovic:2015hza}. It would be interesting to see what is the actual relevant discretization of the simplicity constraint for the \textit{BFCG} case, we leave this question for the future. 

\medskip 

The Euclidean \textit{BF} theory or the \textit{BFCG} theory are equivalent up to a boundary term, so we expect them to have the same physical content. We note that each theory corresponds to different symmetries, both given in terms of 2-groups. In the BF case, we have a pair of trivial  2-groups, where one of the groups is the trivial group, so that the 2-group is really a 1-group (decorating the links or the triangles). In the BFCG case, we deal with the Euclidean 2-group and its dual. At the quantum level, amplitudes can be constructed in terms of the representations of each symmetry structure, be it the Euclidean 1-group or the Euclidean 2-group. Since the two descriptions describe the same theory, this means there must be some relation between the amplitudes expressed in terms of the different representation theories.  Being able to identify explicitly the relation would be extremely interesting.   

\medskip 

In fact some ideas on how to proceed could rely on the concept of semi-dualization \cite{Majid:1996kd, Majid:2008iz}. The  algebraic structure underlying the discretized BF structure, the cotangent bundle $T^*G$, is the Drinfeld double $\fd\sim \iso(4)\semil \iso(4)^*\sim (\R^4\semir \so(4))\semil (\R^{4*}\semir \so(4)*)$ as a Lie algebra. To get to the BFCG formulation, we swapped the sector  $\R^4$ and $\R^{4*}$, which amounts to slicing $\fd$ in a different manner. Still as a Lie algebra, the relevant Drinfeld double for the BFCG formulation is still $\fd$ but now given by 
\begin{align}
    \fd\sim  (\R^{4*}\semir \so(4))\semil (\R^{4}\semir \so(4)^*).
\end{align}
Let us recall quickly what is the  semi-dualization. If we consider a pair of Lie algebras $\g_i$ (equipped with some cocycle), acting on each other so that we can consider the big Lie algebra $\g_1\bowtie \g_2$, then roughly the semi-dualization is defined via the map 
\begin{align}
    \g_1\bowtie \g_2 \rightarrow \g_1\rhd\!\!\!\blacktriangleleft \g_2^*,
\end{align}
where $\blacktriangleleft$ encodes a co-action.   However since we are dealing still with an abelian group $\R^4$, there is no kick-back action and no co-action upon semi-dualization. Note however that the non trivial 2-group appeared thanks to the semi-dualization!  We expect that a deformation of $\R^4$ into a non abelian group such as $AN(3)$ for example would lead to non trivial interesting structures related to Majid's bicrossproduct construction \cite{Majid:1996kd}.

\medskip 
This later consideration brings us to a natural extension of the current work to include the cosmological constant. Instead of working with $\ISO(4)$, we could work with a $\SO(4,1)$ \textit{BF} theory. Under the semi-dualization we would expect to recover the Poincar\'e 2-group decorating the dual complex and a non-trivial 2-group on the triangulation given by the crossed module $AN\ltimes \R^6$, which has a trivial $t$ map. Interestingly to be consistent with the Poisson structure, these 2-groups should have some non-trivial Poisson structure (inherited from the coaction we just mentioned) so that they would be really quantum 2-groups under quantization. Preliminary studies point to the apparition of the $\kappa$-Poincar\'e group and the $\kappa$-Poincar\'e algebra. This is work in progress \cite{wip}.  

\medskip

During our discussion, we did not consider excitations of any kind. We could decorate the vertices with 2-curvature excitation, or the edges with 1-curvature excitation. While a 4d BF theory is naturally coupled with string like excitations \cite{Baez:2006sa,Baez:2006un} it would be interesting to see how these are expressed when using the non-trivial 2-gauge picture, ie the BFCG formulation. We leave this for later investigations.   
\medskip

Finally, we have discussed that BF theory has two main discretizations, the standard one and the dual one where we swap the full group with the dual abelian group. We argued that we can do a partial dualization where we swap only the translational sector to recover the BFCG discretization. There is therefore a fourth case, where we do a partial dualization on the Lorentz sector. This would amount to a ``dual" BFCG discretization. We leave this case for later study.

\section*{Acknowledgment} We would like to thank B. Dittrich and A. Riello for some discussions and suggestions.

\appendix
\section{Charge algebras}\label{algebra}
\subsection{BF case}
The BF charge algebra is 
\begin{align}
    \{ \cP_{\beta'} , \cP_{\beta} \} \coloneqq& \delta_{\beta}\lrcorner (\delta _{\beta'}\lrcorner \Omega) = -\delta_\beta \lrcorner \delta \cP_{\beta'} \\
    =& 0\\
    \{ \J_{\alpha} , \J_{\alpha'}\} \coloneqq& \delta_\alpha \lrcorner (\delta_{\alpha'}\lrcorner \Omega) = \int_M d\la\B \wedge [\alpha', \alpha]\ra-\int_M \la d_\A \B \wedge [\alpha',\alpha]\ra\\
    =& \J_{[\alpha',\alpha]}\\
    \{\J_{\alpha}, \cP_\beta\} \coloneqq& \delta_\alpha \lrcorner (\delta_\beta \lrcorner \Omega) = \int_M d\la [\alpha, \beta]\wedge \A \ra +\int_M \la [\alpha,\beta]\wedge \F \ra+\int_M \la d\alpha\wedge d\beta \ra\\
    =&  \cP_{[\beta,\alpha]}+\int_M \la d\alpha\wedge d\beta \ra
\end{align}
The brackets form a closed algebra up to the extra boundary term in the $\J$, $\cP$ bracket which can be viewed as a central extension. In the following we will focus on parameters $(\alpha,\beta)$ which are constant on the boundary of $M$, so that this central extension vanishes. When picking such parameters we will call the charges,  \textit{global} charges.  

\subsection{BFCG case}
We note that these charges follow directly from decomposing the fields in the charges of BF theory into the subalgebra components. The resulting Poisson brackets are
\begin{align}
    \{ \cR_{Y}, \cR_{Y'}\} = 0, &\quad&  
    \{ \K_X , \cR_{\zeta}\}=0, &\quad& 
    \{ \Q_{\zeta}, \cR_{\zeta}\} =0\\ 
    \{ \cL_\alpha , \K_X\} = \K_{[X,\alpha]}, &\quad& 
    \{ \K_X,\K_{X'}\} = 0, &\quad& 
    \{\cL_\alpha, \Q_{\zeta} \}= \Q_{[\zeta,\alpha]}\\
    \{\Q_{\zeta}, \Q_{\zeta '}\} = 0&\quad& 
    \{\cL_\alpha,\cL_{\alpha'}\} = \cL_{[\alpha',\alpha]} 
\end{align}
We find again the central extension
\begin{align}
  \{\Q_{\zeta}, \K_X\} = \cR_{[X,\zeta] } -\int_M \la d\zeta \wedge dX\ra, \quad
    \{ \cL_\alpha , \cR_{Y}\}=\cR_{[Y,\alpha]} +\int_M \la dY \wedge d \alpha\ra .   
\end{align}
  The Poisson brackets form a closed algebra.

\bibliographystyle{hplain}
\bibliography{biblio1}
\end{document}